\documentclass[twocolumn]{article}

\usepackage[english]{babel}

\usepackage[letterpaper,top=1in,bottom=1in,left=1in,right=1in,marginparwidth=1.75cm]{geometry}

\usepackage[square,numbers]{natbib}
\usepackage{amsmath}
\usepackage{graphicx}
\usepackage[colorlinks=true, allcolors=blue]{hyperref}
\usepackage{caption}
\usepackage{subcaption}
\usepackage{multicol}

\title{Chaos-induced transport due to electrostatic-fluctuations in a toroidal plasma}
\author{Andres E. Medrano-Albarran* and Julio J. Martinell\\Instituto de Ciencias Nucleares, UNAM,
Mexico City, Mexico.\\
*Current Address: Institute for Fusion Studies, UT Austin, Texas, USA.}

\begin{document}
\maketitle

\begin{abstract}
Plasma transport produced by electrostatic fluctuations is studied using a test particle model in a toroidal geometry. 
A model wave spectrum is assumed which introduces a time dependence in the Hamiltonian description of particle motion in the strong magnetic field. This allows for the existence of chaotic particle motion which gives rise to transport processes. Another way of producing transport is by introducing a stochastic process inside the wave spectrum. The two possibilities are explored and it is shown that both cases give rise to diffusive transport. The presence of chaotic motion is corroborated by computing approximate Lyapunov exponents. Additionally the presence of zonal flow is considered which can give rise to a transport barrier by suppressing radial particle fluxes, in the cases that show chaotic transport. This effect gives support to the existence of chaotic motion as responsible for transport since it reproduces the behavior of the robust invariant torus resilience to chaos, acting as a transport barrier, observed in Hamiltonian models of chaos.
\end{abstract}

\section{Introduction}

Turbulent transport is one of the most important problems in magnetic confinement fusion research. It has been studied using different approaches and under a variety of conditions, which give a plethora of different results. As is well known, in magnetically confined plasmas the most common turbulent processes are due to drift waves \cite{horton99}.  
The most common types of drift wave turbulence in toroidal plasmas are due to ion temperature gradient (ITG), trapped electron mode (TEM), and electron temperature gradient (ETG), which drive heat and particle radial fluxes. The detailed study of the drift wave instabilities that give rise to turbulence has to be done with multiscale analyses which require simulations with numerical codes \cite{maeyama24}.

As an alternative to describe the transport, it is possible to use simplified models that can give insights into the physical mechanisms involved. A class of such models deals with the behavior of test particles under the influence of a spectrum of electrostatic waves. Simplified descriptions use a uniform magnetic field and the motion of particles is followed in the two dimensions normal to the field. Adopting a guiding center description, the particle motion is governed by the $E\times B$ drift velocity and the dynamics reduces to a Hamiltonian system \cite{del-castillo93, kleva84}. The wave electric field introduces a time dependence of the Hamiltonian which leads to a non-integrable system. This means that particle trajectories become chaotic when the wave amplitude is large enough in some specific regions of phase-space. In the limit of global chaos the displacements of an ensemble of particles can be described by a transport process.

With this description it has been possible to interpret some observed features of plasma transport as known properties of dynamical Hamiltonian systems. The chaotic orbits appearing for large enough perturbation amplitude are identified as the particle scattering by turbulence due to microinstabilites. They are visualized in Poincaré sections in phase-space. The stochastic-like behavior of a population of particles gives rise to a diffusive transport. Depending on the kind of fluctuations, the diffusion coefficient, $D$, may scale differently with the fluctuation amplitude, $\delta \phi$. When the fluctuations spectrum consists of just two waves $D$ scales linearly with the amplitude but becomes independent of $\delta \phi$ for large amplitudes \cite{kleva84}. For an infinite spectrum, the scaling is $D\sim \delta\phi^2$, as in quasilinear theory \cite{kleva84, kryukov2018, torresTransporteTurbulento2023}.

Corrections to the guiding center approximation have been studied by including finite Larmor radius effects, which are relevant for high energy particles (e.g. alpha particles or NBI ions). Equations are then averaged over the gyroradius which has the effect of chaos reduction \cite{delcastillo-martinell2012, Martinell-dcn2013, kryukov2018, torresTransporteTurbulento2023}. Also, for a thermal distribution of gyroradii the particle distribution functions are no longer Gaussian and acquire long tails, indicating the presence of non-local processes \cite{kryukov2018,torresTransporteTurbulento2023}.

The presence of sheared zonal flows can also be studied with these models. Adding a zonal electric potential to the wave spectrum, the phase-space acquires a structure of nested torii corresponding to a non-twist Hamiltonian system \cite{morrison2000}, meaning that the variation of the winding number from torus to torus (shear) is not monotonic. In this case, chaos starts to develop around torii with rational winding number. As the wave amplitude increases the chaotic region grows until just one torus remains unbroken. This last torus separates two chaotic regions and is related to the shearless surface. It thus acts like a transport barrier \cite{del-castillo93, delCastilloNegrete2000chaotic, delcastillo96, Martinell-dcn2013, ferro_caldas18}. In this way, the presence of transport barriers observed in fusion plasmas with zonal flows \cite{burrell97, 
kobayashi20, macha23} can be explained with two-dimensional Hamiltonian models.

For more realistic situations it would be necessary to extend the particle motion space to three dimensions in a toroidal geometry. Since the dynamics is still governed by a Hamiltonian system with time dependence it is expected that the particle orbits can become chaotic for certain conditions. Under these circumstances the particles would experience stochastic transport similar to what is obtained in 2D models. In order to explore the appearance of chaotic orbits in a toroidal 3D system, we here study the particle dynamics in the guiding center approximation, in the toroidal magnetic field of a tokamak with circular cross section in presence of a drift wave spectrum. This analysis is considerably more complex that the 2D analysis since the representation in phase-space with Poincaré sections to visualize the onset of chaos is not directly possible because of the higher dimensionality, so an alternative approach has to be taken. By directly plotting the particle orbits in a poloidal cross section it is difficult to make out when an orbit becomes chaotic (although a few cases may be identified). Therefore, the approach we take is to analyze an ensemble of particles starting on a single magnetic surface and follow their evolution statistically. We show that the variance of the particle distribution function (PDF) increases with time in some situations which we interpret as the manifestation of transport due to the onset of chaotic orbits. This is reinforced by computing the Lyapunov exponents of the particle trajectories for the cases suspected to present chaos.

The behavior of chaotic orbits as the cause of transport is then compared with another way of driving stochastic transport, namely by introducing random electrostatic fluctuations. This is done by means of random phases in the drift wave spectrum. This will naturally give rise to stochastic orbits and in turn a diffusive transport of a particle ensemble.  The properties of the two types of transport are contrasted in order to show that only the chaotic transport complies with the characteristics of the 2D models. In particular, when the presence of a zonal flow is added, the appearance of a transport barrier can be observed with the same features of the shearless barrier seen in two-dimensional models, only in the case of chaotic transport.

The remaining of the paper is organized as follows. In Section \ref{s2} the particle orbits in the toroidal geometry are described, setting the scenario for a test particle analysis of transport. Section \ref{s3} presents the behavior of the PDF as a way to identify the appearance of transport, both for deterministic and random wave phases. Then, in Section \ref{s4} the computation of Lyapunov exponents for the suspected chaotic orbits is developed. An interesting case in which the particle ensemble separates in two populations is shown is Section \ref{s5} as well as its interpretation. The effect of zonal flows is explored in Section \ref{s6} showing that the establishment of a transport barrier is associated with the presence of chaos. Finally, in Sections \ref{s7} and \ref{s8} the discussion of results and the conclusions are given.

\section{Test particle analysis \label{s2}}

The model used here considers a magnetic field for a toroidal device with a circular cross section.
A test particle is followed in this field according to the guiding center equations in the toroidal geometry, given by White \cite{white2014}. 
%\begin{equation}
%	\begin{aligned}
%		\dot{\psi}_p & =-\frac{g}{D}\left[\left(\mu+\rho_{\|}^2 B\right) \dfrac{\partial B }{\partial \theta} +\dfrac{\partial \Phi }{\partial \theta} \right]\\
%        &\quad +\frac{I}{D}\left[\left(\mu+\rho_{\|}^2 B\right) \dfrac{\partial B}{\partial \zeta} +\dfrac{\partial\Phi }{\partial \zeta} \right] \\
%		\dot{\theta} & =\frac{\rho_{\|} B^2}{D}\left(1-\rho_{\|} g^{\prime}\right)+\frac{g}{D}\left[\left(\mu+\rho_{\|}^2 B\right) \dfrac{\partial B }{\partial \psi_p} +\dfrac{\partial \Phi }{\partial \psi_p} \right] \\
%		\dot{\zeta} & =\frac{\rho_{\|} B^2}{D}\left(q+\rho_{\|} I^{\prime}\right)-\frac{I}{D}\left[\left(\mu+\rho_{\|}^2 B\right) \dfrac{\partial B }{\partial \psi_p} +\dfrac{\partial \Phi }{\partial \psi_p} \right] \\
%		\dot{\rho}_{\|} &=  -\frac{1}{D}\left(1-\rho_{\|} g^{\prime}\right)\left[\left(\mu+\rho_{\|}^2 B\right) \dfrac{\partial B }{\partial \theta} +\dfrac{\partial \Phi }{\partial \theta} \right]\\
%        & \quad -\frac{1}{D}\left(q+\rho_{\|} I^{\prime}\right)\left[\left(\mu+\rho_{\|}^2 B\right) \dfrac{\partial B}{\partial \zeta} +\dfrac{\partial \Phi }{\partial \zeta} \right]
%	\end{aligned}
%	\label{eq:EcuacionesGCM}
%\end{equation}
They are formulated in terms of the magnetic coordinates (the poloidal magnetic flux $\psi_p \sim r^2$, and the poloidal and toroidal angles $\theta, \zeta$), the normalized parallel velocity $\rho_{\parallel} = \frac{v_\parallel}{B}$ and the poloidal and toroidal currents, $g$, $I$.
The equations are derived from a Hamiltonian which is time independent and therefore the trajectories are deterministic. 
Then, the presence of electrostatic drift waves is introduced through a superposition of electric fluctuations representing an spectral distribution. The potential describing the fluctuations is time dependent, and therefore the Hamiltonian system is no longer integrable. This may lead to chaotic behavior of particle trajectories.

\subsection{Magnetic Field Model}

The magnetic field has toroidal and poloidal components $
\vec{B}(r, \theta) = B_{\zeta}(r, \theta)\hat{\zeta} + B_{\theta}(r, \theta)\hat{\theta}$. The toroidal field for circular flux surfaces, having the $1/R$ dependence, is simply given by

\begin{equation}
	B_{\zeta}(r, \theta) = \dfrac{B_0}{1 + \frac{r}{R_0}\cos(\theta)}, 
	\label{eq:ModMagMartinell}
\end{equation}
where $B_0$ is the magnetic field at the magnetic axis. The poloidal field is given in terms of the $q$-factor as $B_\theta= rB_\zeta/R_0 q(r)$. For $q(r)$, it is assumed to have a parabolic profile, as it is commonly the case for tokamaks \cite{wessonTokamaks2004},

\begin{equation}
     q(r, \theta) = q_0\left(1+ \dfrac{r^2}{\lambda^2}\right),
\end{equation}
where $q_0$ is the safety factor at the axis and  $\lambda^2 = \dfrac{a^2 q_0}{q_0-q_w}$ with $q_w$ the safety factor at the edge.

This model is assumed to correspond to the equilibrium magnetic field and thus it is held fixed for all calculations. The {\it test} particles that move in this field do not modify it.

\subsection{Wave spectrum}

The electrostatic potential for the drift waves is taken to be a superposition of sinusoidal waves propagating in the poloidal direction of the same amplitude,

\begin{equation}
	\phi(r,\theta, t) = A \operatorname{sech}^2 \left(\frac{\frac{r}{a} -c }{\Delta}\right) \sum_{j=1}^{nw}\operatorname*{sin}\left(\theta - j t +\varphi_{j}\right),
	\label{eq:PotencialFinal}
\end{equation}
There is a radial modulation that introduces a localization of the wave spectrum. This also reduces the effect of the waves when particles approach the walls, which are not included in the simulations. The time dependence of $\phi$ is what can make the Hamiltonian system to be non-integrable. For the phases $\varphi_{j}$, two different cases are considered: (a) they have deterministic values (given by $\varphi_j=2\pi/j$) or (b) they are taken from a random distribution. The first case is relevant for the exploration of a chaotic behavior, while the second produces a stochastic process. The number of waves that forms the spectrum $nw$, is taken as a variable.  

%\begin{figure}
%\centering
%\includegraphics[width=0.25\linewidth]{pdf}
%\caption{\label{fig:frog}This frog was uploaded via the file-tree menu.}
%\end{figure}

\subsection{Types of orbits}

The first step in our a analysis is to characterize the particle orbits moving in the magnetic and electric fields described above.  Depending on the initial pitch angle the trajectories in a toroidal magnetic field can be passing (circling the whole torus) or trapped (bouncing between magnetic mirror points). When the wave potential is introduced they can be modified through the $\vec{E}\times \vec{B}$ drift. For the potential given by Eq.~\ref{eq:PotencialFinal} the electric field has poloidal and radial components although the relevant one is the former. Thus, the electric drift produces a displacement in the radial direction and rotations in the poloidal and toroidal directions given by 
$\vec{v}_E = \dfrac{1}{B^2}\left[  E_{\theta} B_{\zeta} \hat{r}  - E_{r} B_{\zeta} \hat{\theta} + E_{r} B_{\theta} \hat{\zeta} \right]$
where the angular components are subdominant.

When the initial pitch angle corresponds to a passing orbit, due to the poloidal symmetry of the trajectory, the only discernible effect of the electric potential is a radial drift, increasing the orbit radius. In contrast, for a trapped particle orbit, both radial and poloidal drifts can be observed in a poloidal projection.

\begin{figure}[htp]
	\centering
	\begin{subfigure}[t]{0.49\textwidth}
		\centering
		\includegraphics[width=0.85\linewidth]{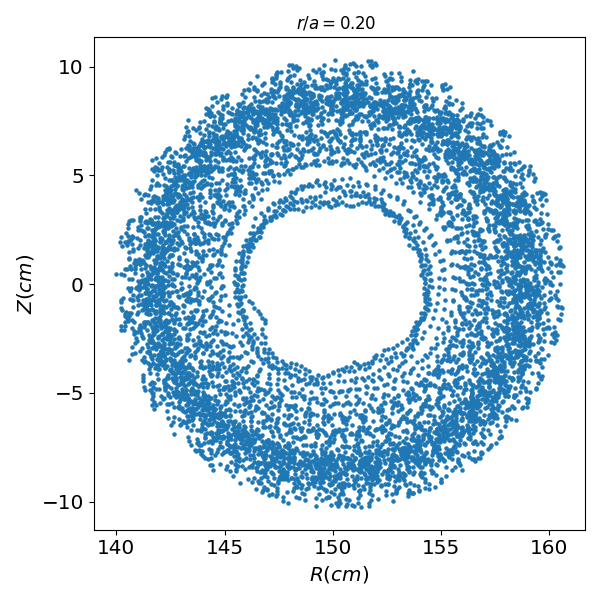}
		\caption{Drift over a passing particle.}
	\end{subfigure}
	\vspace{-0.1cm}
    
	\begin{subfigure}[t]{0.49\textwidth}
		\centering
		\includegraphics[width=0.85\linewidth]{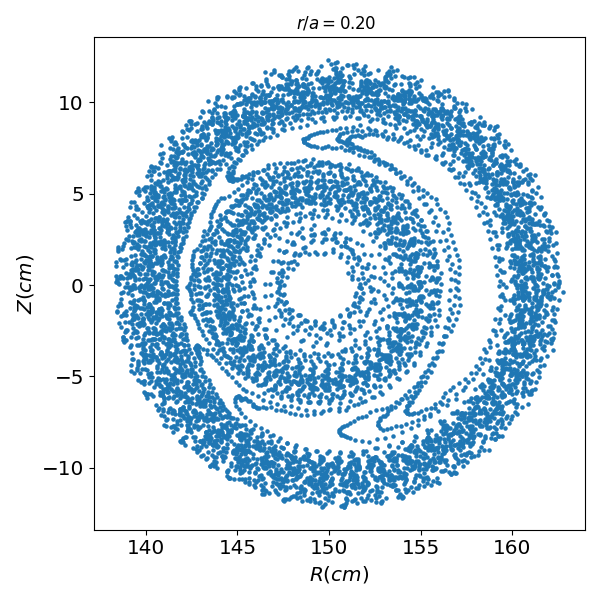}
		\caption{Drift over a trapped particle.}
	\end{subfigure}
	
	\vspace{-0.3cm}
	
	\caption{Poincaré plots of particle orbits on a poloidal plane showing the effect of the electric drift induced by an electric potential amplitude of $A = 0.1\mathrm{V}$, 10 waves and {\it random} phases. (a) Passing particle affected only by the radial drift, increasing the radius. (b) Trapped particle affected by both radial and poloidal drifts, transforming the orbit into a passing one.}
	\label{fig:ORBITS_DRIFTS_RDM}
	
\end{figure}

For the case of random phases, the orbit of a passing particle spans a radial width since the radial drift is changing between positive and negative values as it can be seen in the poloidal Poincaré plot of Figure \ref{fig:ORBITS_DRIFTS_RDM}(a). Trapped particles additionally exhibit the effect of the poloidal drift and in some cases the combined action of these drifts transforms a trapped orbit into a passing one, as illustrated in Figure \ref{fig:ORBITS_DRIFTS_RDM}(b). This transition results from the energy transferred to the particle by the electric potential, which modifies its pitch angle and consequently changes the relative contributions of the parallel and perpendicular velocity components.

\begin{figure}[hbp]
	\centering
	\begin{subfigure}[t]{0.49\textwidth}
		\centering
		\includegraphics[width=0.80\linewidth]{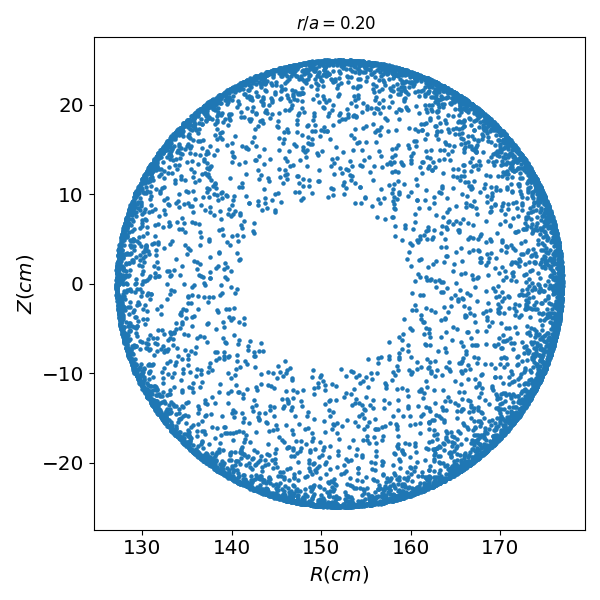}(a)
	%	\caption{Radial drift over a passing particle.}
	\end{subfigure}
	\vspace{-0.1cm}
	
	\begin{subfigure}[t]{0.49\textwidth}
		\centering
		\includegraphics[width=0.8\linewidth]{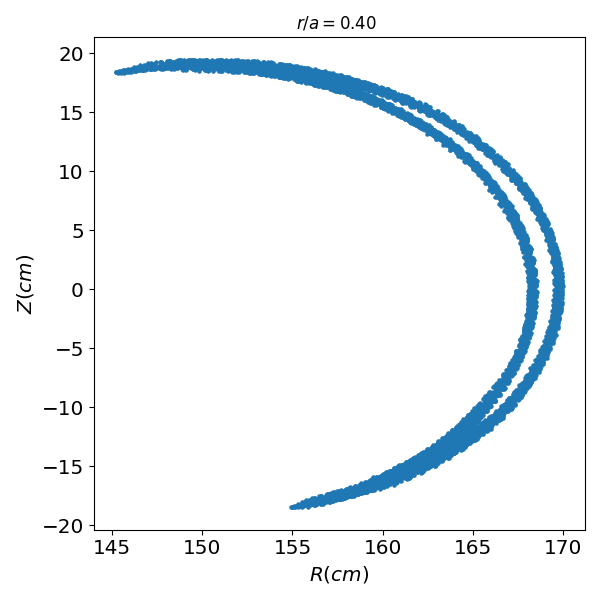}(b)
	%	\caption{Full drift over a trapped particle.}
	\end{subfigure}
    
	\vspace{-0.1cm}
    
	\begin{subfigure}[t]{0.49\textwidth}
		\centering
		\includegraphics[width=0.80\linewidth]{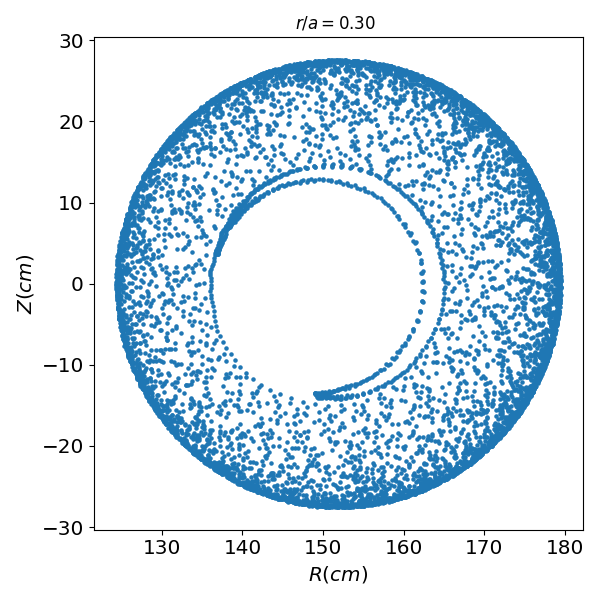}(c)
	%	\caption{Drift on a trapped particle transforming into a passing one.}
	\end{subfigure}
	
	\vspace{-0.3cm}
	
	\caption{Poincaré plots of particle orbits on a poloidal plane showing the effect of the electric drift induced by an electric potential amplitude of $A = 1\mathrm{V}$, 10 waves and {\it deterministic} phases. (a) Radial drift, increasing the radius of a passing orbit. (b) Anticlockwise poloidal drift of a trapped orbit. (c) Combined radial and poloidal drifts, transforming a trapped orbit into a passing one.}
	\label{fig:ORBITS_DRIFTS_DET}
\end{figure}

For deterministic phases, the particle evolution follows the expected deterministic trajectory, although in some cases it may exhibit deterministic chaos because of the time dependence of the electric potential. In Figures \ref{fig:ORBITS_DRIFTS_DET}(a) and (b) the trajectories of passing and trapped particles are shown. The starting point for the passing particle has a small minor radius ($r=10$ cm) and it drifts outwards due to the combined $E\times B$ and grad-B drift velocities. For the trapped particle, it shows the poloidal drift displacement together with a small radial widening. Meanwhile, in Figure \ref{fig:ORBITS_DRIFTS_DET}(c) a transition from trapped to passing orbit is depicted, which occurs for pitch angles of barely trapped orbits.

Although chaotic orbits are expected to appear for certain initial conditions and perturbations, it is not clear a priori how to determine their appearance. A systematic study of the transition to chaos is not possible in this tridimensional system in the same way as it is done in two dimensional studies like \cite{kleva84, kryukov2018, del-castillo93, Martinell-dcn2013}. Chaotic orbits can be identified in some cases but in general it is not possible to assure the chaotic nature of an orbit. 

That is why, we chose to turn to a different method to identify the presence of chaotic orbits. 
For this, an ensemble of particles is released in order to determine the ensuing evolution which gives rise to a particle distribution that evolves in time. They are initialized with almost the same conditions so that the evolution of the distribution function can inform about the possible presence of chaos.

\section{Distribution functions \label{s3}}

\begin{figure}[htp]
	\centering
		\includegraphics[width=0.85\linewidth]{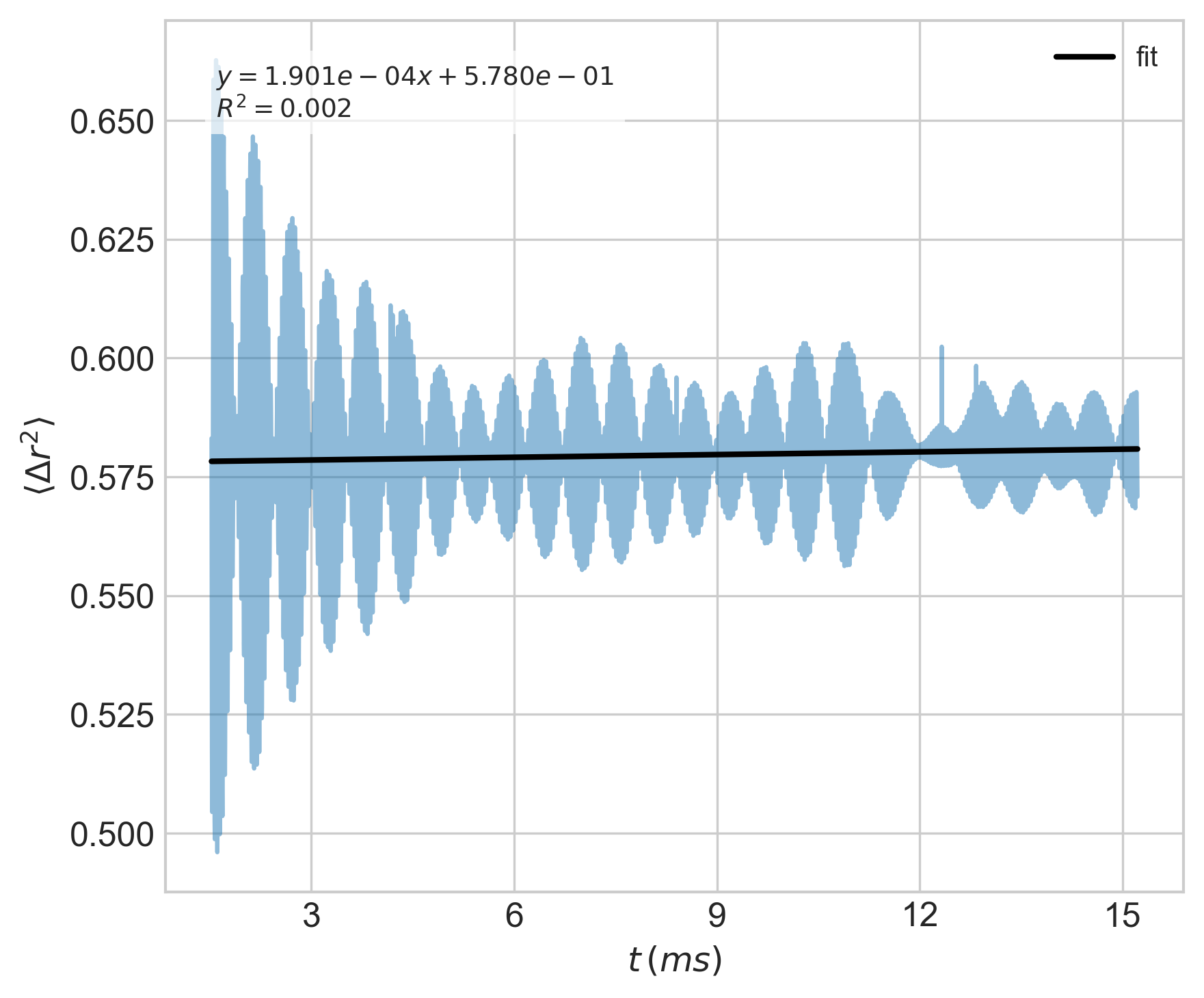}
	\caption{PDF variance evolution shows that in absence of electric drift, even though the variance oscillates due to the grad-B and curvature drifts, it does not increase with time.}
	\label{fig:A0_R2_E1_n2_PT20_fix_SUPER}
\end{figure}

Since it is not easy to determine when and if the orbits become chaotic, it is more convenient to follow an ensemble of particles released in a certain region of the plasma. They can evolve in a way that would indicate the presence of chaotic orbits and transport. For instance, if a particle ensemble is initially distributed on a single magnetic surface, the expected evolution (if there is no chaos) would only lead to the displacements due to the guiding center drift velocities, namely, grad-B and curvature (see Figure \ref{fig:A0_R2_E1_n2_PT20_fix_SUPER}). But the presence of chaos should lead to diverging particle trajectories that move particles away from the initial magnetic surface in all directions, giving rise to a widening of the radial distribution function.

In the simulations that follow the electric potential for the fluctuations is given by Eq.~\ref{eq:PotencialFinal} with $c=0.5$ and $\Delta=0.4$. The amplitude $A$ and the number of waves in the spectrum $nw$ are varied in the ranges $0.01\le A\le 1$ V and $1\le nw \le 20$. For the particle population, the initial radial position is varied in the range $0.2\le r_i/a \le 0.8$, while the pitch angles are fixed at $20^\circ$ (circulating orbit) or $70^\circ$ (trapped orbit); in some cases the distribution is taken isotropic (uniform pitch angle) which will be noted. 
The two cases of random and deterministic phases of the waves will be analyzed to study the distribution functions.

% \subsection{Transport model.}

% NOTE: I propose to add this section in order  to describe the model for transport and of course the refences, maybe something like the analysis made in kryukov \cite{kryukov2018}.

% To study the transport produced by the chaos or the stochastic motion, it is natural to use a diffusive model,

% \begin{equation}
% 	\sigma(t) = \alpha t^{\beta},
% 	\label{eq:ModTrans}
% \end{equation}

% which tell us how the variance of the distribution is evolving through time, and hence how is the width of the distribution changing. Here $\alpha$ is the diffusion coefficient and $\beta$ is the diffusion exponent.  Usually, $\beta = 1$ corresponds to normal diffusion, $1 < \beta$ to superdiffusion, and $\beta < 1$ to subdiffusion, allowing the transport regime to be classified.

\subsection{Random phases \label{s3r}}

In order to have an alternative transport process due to the presence of waves, against which to compare the hypothesis of a chaotic transport, we consider a stochastic process where the phases are chosen from a random distribution.

\begin{figure}[htp]
	\centering
	\begin{subfigure}[t]{0.48\textwidth}
		\centering
		\includegraphics[width=0.85\linewidth]{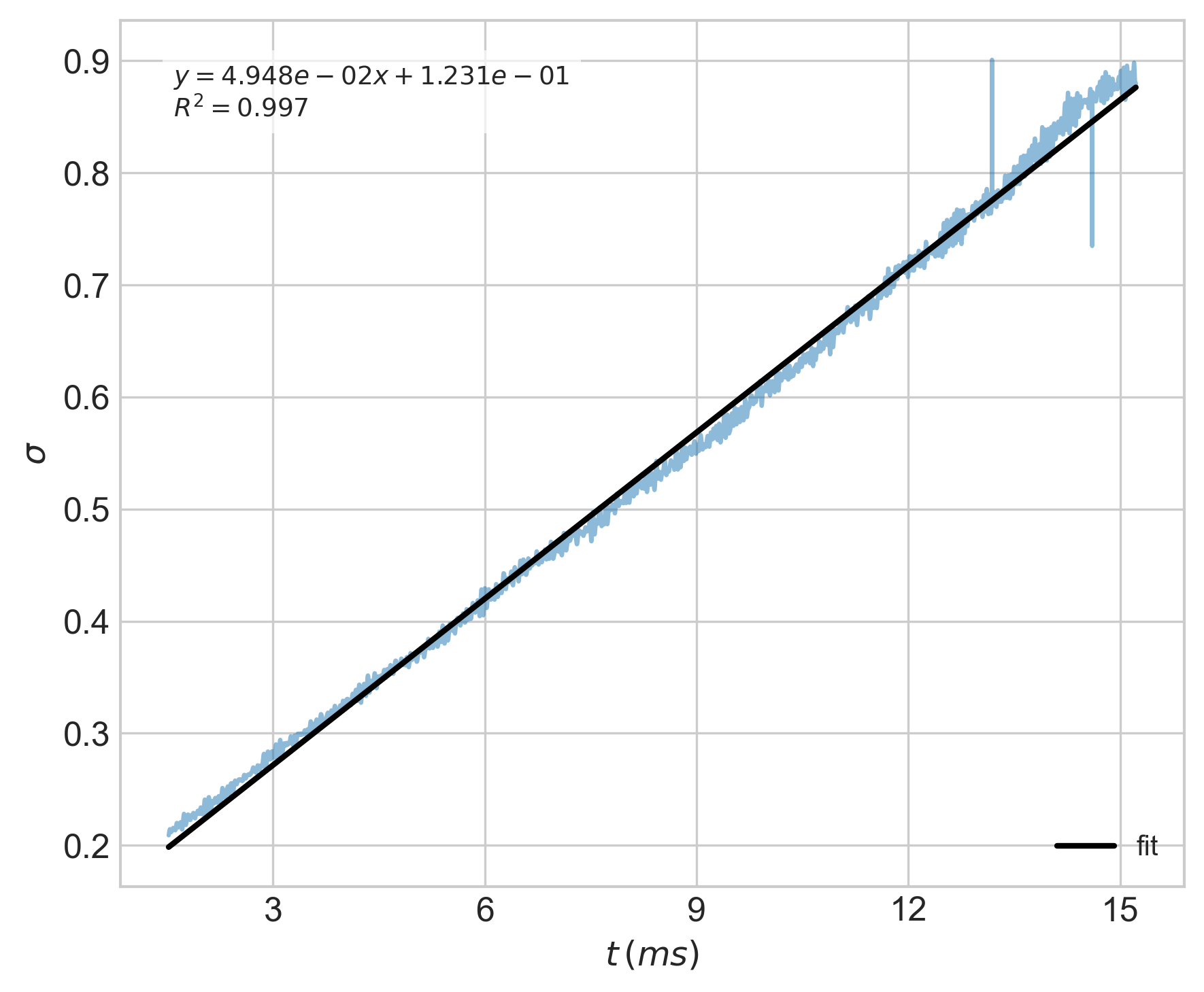}
		\caption{PDF Variance evolution.}
	\end{subfigure}
    	\begin{subfigure}[t]{0.48\textwidth}
		\centering
		\includegraphics[width=0.85\linewidth]{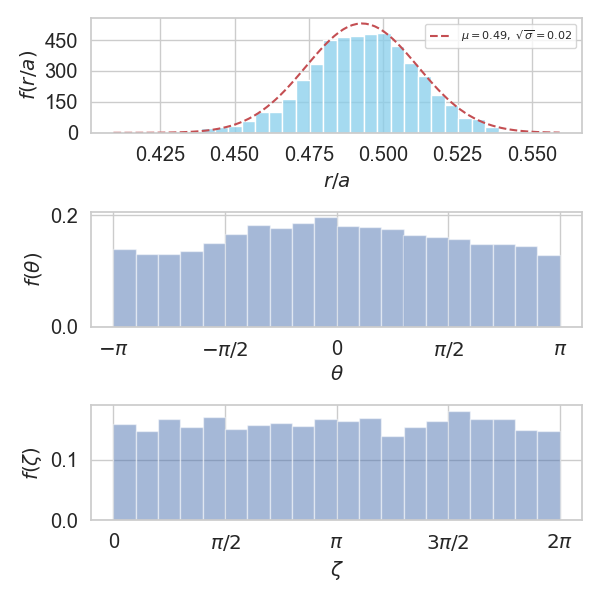}
		\caption{Final distribution.}
	\end{subfigure}

	\caption{Particle distribution evolution for the case $A=0.1$ V and $nw=2$ waves, with random phases considering initial $\frac{r}{a} =0.5$ and $\alpha = 20^{\circ}$.
    (a) Shows a linear dependence of the variance with time, indicating the presence of diffusive transport. (b) Shows that the final PDF is a gaussian function in radius while the angular distributions are uniform.}
	\label{fig:A10e4_R5_E1_n2_PT20_fix_RDM}
\end{figure}

The general behavior exhibits that when the effective amplitude of the potential is large enough to significantly modify the deterministic trajectories due to the magnetic field, transport can be observed.  Here we identify transport as the presence of a widening of the radial distribution, which is observed as an average increase of the PDF variance in time while maintaining an approximate gaussian shape. All simulations start with a particle ensemble at a single magnetic surface for a given radius (which was changed for different simulations) and uniformly distributed in toroidal and poloidal angles. In all of the reported cases there was a linear evolution of the variance with time ($\sigma=\langle \Delta r^2 \rangle\sim t$), which corresponds to a normal diffusion process, as expected. Figure \ref{fig:A10e4_R5_E1_n2_PT20_fix_RDM} illustrates both indicators associated with transport for a particular case: in (a) there is a clear linear increase of $\sigma$ with time, while in (b) the final distribution function has a gaussian shape in radius and it is uniform in poloidal and toroidal angles. Notice that the variance is computed by fitting a gaussian to the PDF at different time intervals in the evolution.

This same general behavior is observed for different conditions changing the wave amplitudes, number of waves and initial radial positions. This indicates that transport is present whenever there is a stochastic factor in the wave spectrum, here represented by the random phases of the waves.

The additional factor relevant in the simulations is the initial pitch angle. This determines whether the particles are trapped or passing. Cases with iso-pitch and mono-pitch angles were tested. In the later case it was found that when the pitch angle is small, corresponding to initially passing trajectories, 52 out of 200 simulations with different initial conditions (varying $A, nw$ and starting $r/a$) presented transport. In contrast, for large pitch angle for which the particles are all trapped initially, only 39 out of 200 initial conditions presented transport. This is due to the fact that trapped particles do not present a net radial displacement and thus there is no radial transport; only when they become de-trapped they exhibit transport (see Fig.~\ref{fig:ORBITS_DRIFTS_RDM}(b)). The existence or not of transport is linked to the effective action of the fluctuations; for instance, for small $A$ the effect is too weak and related to this, for radial distances far from $r/a=c$ the intensity of the waves is reduced due to the radial modulation given by the $\operatorname{sech}^2$ term in Eq.~\ref{eq:PotencialFinal}.

\begin{figure}[htp]
	\centering
%	\begin{subfigure}[t]{0.48\textwidth}
%		\centering
		\includegraphics[width=0.95\linewidth]{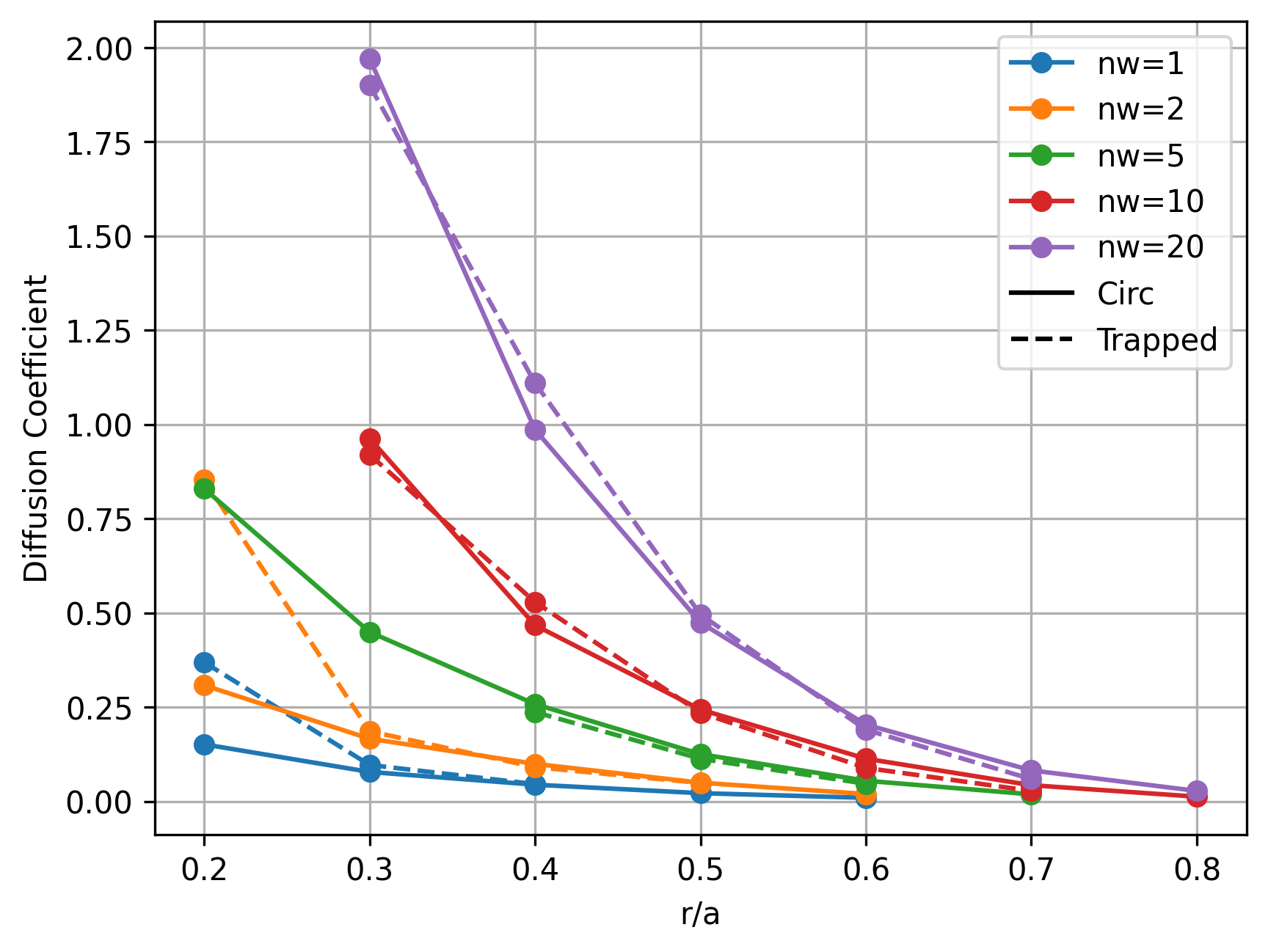}
%        \caption{Diffusion Coefficient vs number of overlapping waves.}
%	\end{subfigure}
%	\begin{subfigure}[t]{0.48\textwidth}
%		\centering
%		\includegraphics[width=0.95\linewidth]{Imagenes/nw_COMB.png}
%		\caption{Diffusion Coefficient vs number of overlapping waves.}
%	\end{subfigure}
	\vspace{-0.15cm}
	\caption{Dependence of the diffusion coefficient with the initial radial position $r/a$ for different number of component waves in the spectrum, for a potential amplitude $A = 0.1$ V. This shows the cases where transport is observed, when the initial pitch angle corresponds to trapped (dashed lines) or circulating particles (solid lines).}
	\label{fig:Slope_RDM}
\end{figure}

The impact of the parameter variation can be quantized in terms of the diffusion coefficient, $D$ ,defined from the slope of the variance time evolution, i.e. $\sigma=D t$. Figure \ref{fig:Slope_RDM} shows the dependence of $D$ with the initial radial position and the number of waves in the spectrum, $nw$. It is clear the increasing the radial position reduces the diffusion coefficient, while it increases with the number of wave components. Thus, the effective amplitude of the potential is related to the magnitude of the diffusion coefficient, since the wave amplitude decreases as the particles approach the plasma edge region. The cases marked as trapped and circulating refer to the initial pitch angle type. 

% \begin{figure}[htbp]
% 	\centering
% 	\begin{subfigure}[t]{0.48\textwidth}
% 		\centering
% 		\includegraphics[width=0.95\linewidth]{Imagenes/ResultadosTRANS/PT70_A105_RDM_ra.png}
%         \caption{Diffusion Coefficient vs number of overlapping waves.}
% 	\end{subfigure}
% 	\begin{subfigure}[t]{0.48\textwidth}
% 		\centering
% 		\includegraphics[width=0.95\linewidth]{Imagenes/ResultadosTRANS/PT70_A105_RDM_nw.png}
% 		\caption{Diffusion Coefficient vs number of overlapping waves.}
% 	\end{subfigure}
% 	\vspace{-0.15cm}
% 	\caption{Changes in the diffusion coefficient  for the potential amplitude A = 0.1 V. }
% 	\label{fig:A10e4_EVO_SLOPE_PT70_RDM}
% \end{figure}

% \begin{figure}[htbp]
% 	\centering
% 	\begin{subfigure}[t]{0.48\textwidth}
% 		\centering
% 		\includegraphics[width=0.95\linewidth]{Imagenes/ResultadosTRANS/PT20_A105_RDM_ra.png}
% 		\caption{Diffusion Coefficient vs radial position.}
% 	\end{subfigure}
% 	\begin{subfigure}[t]{0.48\textwidth}
% 		\centering
% 		\includegraphics[width=0.95\linewidth]{Imagenes/ResultadosTRANS/PT20_A105_RDM_nw.png}
% 		\caption{Diffusion Coefficient vs number of overlapping waves.}
% 	\end{subfigure}
% 	\vspace{-0.15cm}
% 	\caption{Changes in the diffusion coefficient  for the potential amplitude A = 0.1 V. }
% 	\label{fig:A10e4_EVO_SLOPE_PT20_RDM}
% \end{figure}

\subsection{Deterministic phases \label{s3d}}

When the wave phases are not random there is no intrinsic stochasticity and therefore the particles in the ensemble should evolve similarly and no widening of the population would be expected. Thus, any broadening in the PDF would indicate the presence of chaos, as this is the only possibility for a stochastic-like process. As before, cases where particles are initially trapped or circulating (having mono-pitch-angle distribution) were analyzed. In this case, for initially trapped orbits, all potential configurations and number of waves produced no particle transport. This means that the distribution function did not evolve to a Gaussian and no broadening was observed. Although the ensemble experienced the effects of the electric drift, these effects were identical for all particles, preserving an ordered motion and therefore the distribution function width never increased. It can be concluded that, for the potential configurations applied, trapped particle trajectories did not show chaotic behavior within the parameter regime studied.

\begin{figure}[htp]
	\centering
    \begin{subfigure}[t]{0.48\textwidth}
		\centering
		\includegraphics[width=0.75\linewidth]{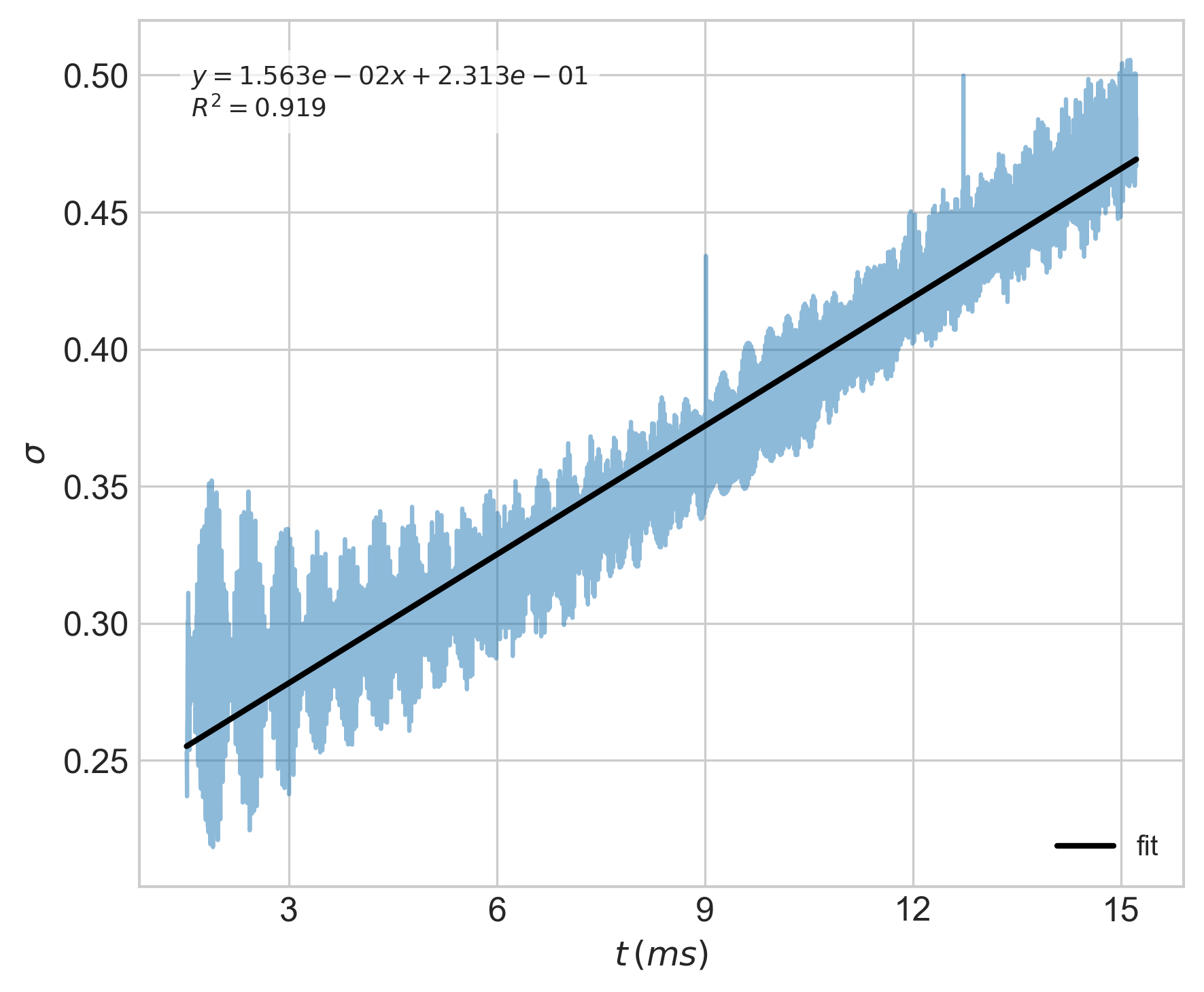}
		\caption{Variance evolution.}
	\end{subfigure}
	\begin{subfigure}[t]{0.45\textwidth}
		\centering
		\includegraphics[width=0.75\linewidth]{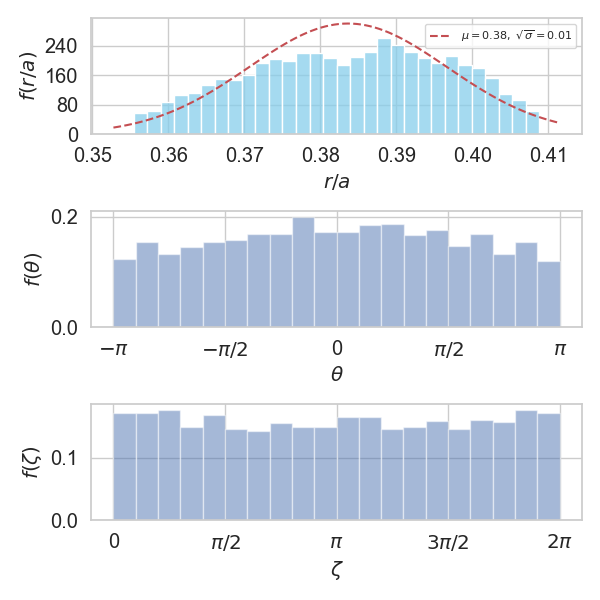}
		\caption{Final distribution.}
	\end{subfigure}
    \caption{PDF for deterministic phases with $A=1V$, one wave, initial radius $r/a=0.3$ and fixed initial pitch of $20^\circ$; (a) time evolution of the radial variance showing an average linear increase, as represented by the black line fit, indicating a widening of the PDF. (b) Displays the final PDF which is maintained as a Gaussian function in radius, while angular distributions are uniform. }
	\label{fig:A10e5_R3_E1_n1_PT20_fix}
\end{figure}

For passing particles, as in the case of random phases, there are cases where the particle distribution function (PDF) variance evolves as a linear function of time which indicates that the wave fluctuations induce a diffusive transport. Table \ref{tab:FD_10E5_PT20} shows the four configurations that presented a widening of the PDF while preserving a gaussian distribution. In figure \ref{fig:A10e5_R3_E1_n1_PT20_fix} it is possible to witness this behavior.

\begin{table}[htp]
	\centering
	\caption{Parameters of the cases that showed transport properties, for a potential amplitude of $A = 1\mathrm{V}$.}
	\label{tab:FD_10E5_PT20}
	
\begin{tabular}{|c c c c |}
	\hline
	$\frac{r}{a}$ & $n_w$ & Diffusion Coefficient & $R^2$ \\
	\hline
		
	0.2 & 2 & $3.60\times10^{-2}$ & 0.94  \\
		
	0.3 & 1 & $1.56\times10^{-2}$ & 0.92  \\
		
	0.4 & 1 & $1.28\times10^{-2}$ & 0.91  \\
	
	0.5 & 1 & $7.31\times10^{-3}$ & 0.81  \\
	
	\hline
\end{tabular}
\end{table}

It is conjectured that the cause of diffusive transport is related to the appearance of chaotic trajectories for the conditions listed in Table \ref{tab:FD_10E5_PT20}. For the other cases the non-integrability conditions for the presence of chaos do not seem to be met. As discussed above, the possibility of chaotic orbits is rooted in the previous results in two dimensional systems \cite{kleva84,del-castillo93,delCastilloNegrete2000chaotic, Martinell-dcn2013} which would be expected to appear in this 3D system. However, since they are not actually seen in the simulations, the chaotic nature of the orbits should be corroborated by other means. One such test is the computation of the Lyapunov exponents which characterize chaos, which is done in the following section.

\section{Lyapunov exponents \label{s4}}

The origin of transport in models with deterministic or constant phases is associated with a chaotic regime that causes each particle in the ensemble to explore different trajectories, ultimately giving rise to a transport phenomenon that is manifested through the widening of the radial distribution. In a two-dimensional model such as that of \cite{kleva84,kryukov2018}, chaos can be detected directly by visualizing the equivalent 1D phase-space, which actually represents all the particle orbits in the domain. In two-degrees-of-freedom systems, periodic particle trajectories are mapped into the phase-space of a single coordinate by plotting Poincaré surfaces of section \cite{Chaos_lichtenberg_lieberman_1983}. Here, chaos is revealed when the surfaces are broken into apparently disordered points. However, when extending to higher-dimensional models, Poincaré sections are not a useful representation. For a toroidal geometry such as that of a tokamak, one can produce some kind of Poincaré plots in a poloidal cross section where, in absence of fluctuations, trapped particles are seen as banana orbits and circulating particles as closed circles. However, when the wave spectrum is introduced the orbits shown in Fig.~\ref{fig:ORBITS_DRIFTS_DET} acquire a radial extent and the appearance of chaos could not be clearly discerned. This would limit the capability of diagnosing chaos.

Several methods exist for detecting the presence of chaos in a more quantitative way, among them the \textbf{Finite-Time Lyapunov Exponent} (FTLE). Given that a chaotic system exhibits sensitive dependence to the initial conditions \cite{strogatz2014nonlinear}, two trajectories that start with very similar initial conditions at time $t_0$ increase their separation exponentially after a time $T$ according to

\begin{equation}
||\delta\vec{x}(t_0+T)|| = ||\delta\vec{x}(t_0)||e^{\lambda T}.
\label{eq:DistanciaLyapunov}
\end{equation}

When the time $T$ tends to infinity, $\lambda$ is defined as the Lyapunov exponent. It effectively measures the growth of the distance between two points with very close initial conditions. It is useful to write it explicitly as

\begin{equation}
\lambda =\lim_{T\rightarrow\infty}  \dfrac{1}{T}\log\left(\dfrac{||\delta\vec{x}(t_0+T)||}{||\delta\vec{x}(t_0)||}\right).
\label{eq:FTLE}
\end{equation}

A positive Lyapunov exponent ($0<\lambda$) leads to an exponential growth in the separation between nearby trajectories. Therefore, it can be used as a criterion for identifying the presence of chaos.

For some dynamical systems, it is possible to obtain this exponent analytically from the geometric nature of their governing equations. However, in most cases it is determined numerically using methods such as those of Benettin or Wolf \cite{Chaos_lichtenberg_lieberman_1983}. These methods compute the evolution of the distance between nearby initial conditions over finite time windows, making it possible to estimate the value the Lyapunov exponent as the value toward which the FTLE eventually converges. All these methods rely heavily on numerical computation and on a rescaling procedure that prevents memory overflow, ultimately yielding a finite time Lyapunov exponent.

Using the concept of the Lyapunov exponent and based on the Wolf method together with reference \cite{Finn-delCastillo2001}, the following diagnostic procedure was applied in order to have more support on a possible chaotic regime in the simulated ensembles and thus explain the origin of the transport phenomenon.

This was done obtaining the trajectories of 5000 particles with nearby initial conditions, meaning that the particles were positioned within an interval of $\pm 10^{-5}$, around the position being tested, in the initial radial position ($\frac{r}{a}$), with equal probability, while having the same angular coordinates, energy, and pitch angle.
Then, their trajectories were followed in time. By considering them in pairs, consisting of the original trajectory and each perturbed one, it was possible to calculate the distance between them at every simulation time. It is important to recall that 1000 integration steps were performed per transit, resulting in an effective time step on the order of microseconds.

By directly applying Eq. \ref{eq:FTLE}, it is possible to obtain a \textbf{finite-time} approximation of the Lyapunov exponent using the elapsed simulation time and the distance between the original and perturbed trajectories. This procedure is repeated for each perturbed trajectory, and the resulting Lyapunov exponents are averaged at every time step to obtain a statistically meaningful value.

It should be noted that, since the system is confined in toroidal coordinates, measuring a distance required unwrapping the angular coordinates using the number of completed turns according to

\begin{equation}
\begin{aligned}
\theta_{U}&= (\theta_0 + 2 \pi N_{\theta})r, \\
\zeta_{U}&= (\zeta_0 + 2 \pi N_{\zeta})(R+r\cos(\theta)).
\end{aligned}
\label{eq:Desdobladas}
\end{equation}

The results shown in table \ref{tab:FD_10E5_PT20} are candidates for transport induced by a chaotic regime, and therefore their finite-time Lyapunov exponents were computed as described above.

Table \ref{tab:FTLE} shows that all cases yielded positive values of $\lambda$, indicating an exponential growth in the distance between trajectories with nearby initial conditions and thus providing evidence of chaos. Furthermore, it can be observed that the response time ($\lambda t \sim 1$) occurs on the order of milliseconds, in agreement with the simulation results, where $\lambda$ is of the order of $10^{3}\mathrm{s}^{-1}$. As a control case, the FTLE was also calculated for a configuration without drift waves, yielding $\lambda\sim0.1\ll10^{3}$, indicating the absence of exponential growth in the separation between trajectories during the simulation time and, consequently, the absence of a chaotic regime, as expected. After an initial transient, the average finite-time Lyapunov exponent approached an approximately constant value. Its asymptotic value was estimated by taking the time average over this plateau region. This behavior was consistently observed in all analyzed cases.

\begin{table}[htp]
\caption{Finite-time Lyapunov exponent approximation for $A=1,\mathrm{V}$.}
\label{tab:FTLE}
\centering
\begin{tabular}{|c c c|}
\hline
$\frac{r}{a}$ & $n_w$ & $\lambda$ $[s^{-1}]$ \\
\hline
0.2 & 2 & 1206.8 \\
0.3 & 1 & 1197.3 \\
0.4 & 1 & 1187.3 \\
0.5 & 1 & 1161.6 \\
\hline
\end{tabular}
\vspace{-0.2cm}

\end{table}

% Figure \ref{fig:a10e5r2e1n2pt20fixsuperftletor} shows the average value of the Lyapunov exponent as a function of time, illustrating that once the transient region is surpassed, the data begin to approach the average asymptote, which was calculated as the mean value of the plateau region. This behavior was consistently observed in all analyzed cases.

% \begin{figure}[hbp]
% \centering
% \includegraphics[width=0.85\linewidth]{Imagenes/A10e5_R2_E1_n2_PT20_fix_SUPER_ftle_tor}
% \vspace{-0.2cm}
% \caption{Evolution of the Finite-Time Lyapunov Exponent for $r/a=0.2$ and $n_w=2$.}
% \label{fig:a10e5r2e1n2pt20fixsuperftletor}
% \end{figure}

Although the obtained values are consistent and exhibit the expected behavior, these simulations provide only a diagnostic that may indicate the presence of chaos. They do not constitute irrefutable evidence that chaos is indeed present, since they depend on the statistical significance of the computed value and on whether convergence has actually been achieved. When the FTLE computed by this method are reasonable, this can give some confidence on the assumption that the wave spectrum system is presenting chaotic behavior.

\section{Separation of populations. \label{s5}}

In considering the cases of deterministic phases, the two situations of mono-pitch-angle and iso-pitch-angle for the initial particle ensemble were included. In the latter case an interesting situation arises. For this case, the spatial initial conditions of the ensemble were identical, but each particle was assigned an equiprobable random value of the pitch angle, which made it possible to obtain multiple cases that presented distribution broadening and a linear variance growth. However, since the initial conditions were not identical, in some situations this apparent broadening resulted from a population separation process rather than the single gaussian population being widened, as shown above. This was apparently the result of having different initial velocity directions of each particle, because the value of the pitch angle determines whether the particle is trapped or passing.

As noted in the previous subsections, the behavior of trapped and passing trajectories differed, with transport effects being reduced or even absent in the presence of trapped trajectories. For this reason, ensembles with random pitch angles naturally split into those that, in the absence of drift waves, would correspond to trapped trajectories and those that would correspond to passing trajectories. Based on this fact, it is hypothesized that the population separation phenomenon represents the separate evolution of trapped and passing populations. Several results exhibited this property.

\begin{figure}[htp]
\centering
\begin{subfigure}[t]{0.45\textwidth}
\centering
\includegraphics[width=0.85\linewidth]{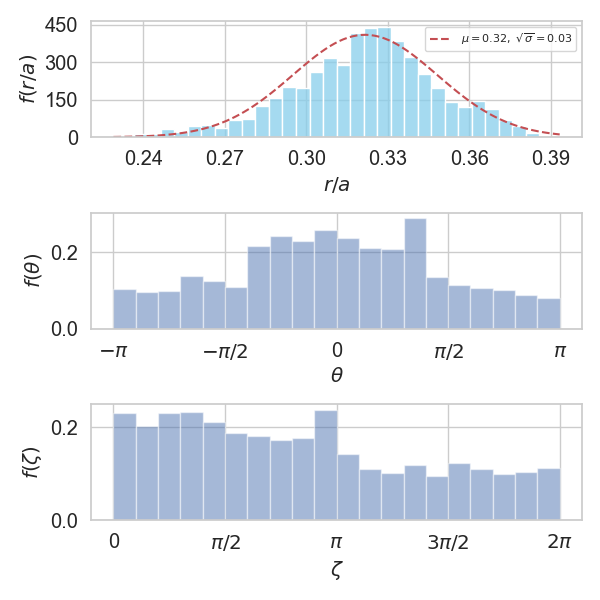}
\caption{Distribution before separation.}
\end{subfigure}
\begin{subfigure}[t]{0.45\textwidth}
\centering
\includegraphics[width=0.86\linewidth]{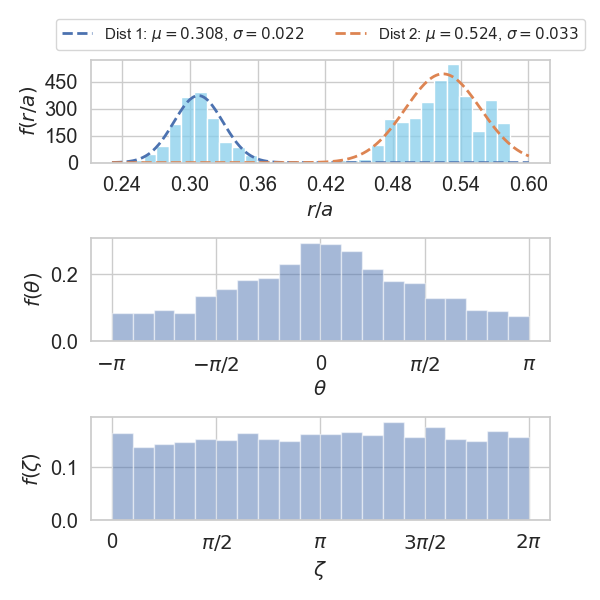}
\caption{Final distribution showing separation into two populations.}
\end{subfigure}
\caption{Distribution functions at two times during the evolution for $r/a=0.3$,  $n_w=5$ and $A=1$ V. Notice that the poloidal distribution in not uniform: the maximum near $\theta=0$ reveals the presence of trapped particles.}
\label{fig:A10e5_R3_E1_n5_PT20_ISO}
\end{figure}

Figure \ref{fig:A10e5_R3_E1_n5_PT20_ISO} illustrates the behavior described above, where, starting from a single distribution at time 0.6 ms, the distribution gradually separates into two populations as it evolves up to 15.22 ms. These populations are hypothesized to correspond to trapped and passing trajectories.

\begin{figure}[htp]
\centering
\begin{subfigure}[t]{0.45\textwidth}
\centering
\includegraphics[width=0.95\linewidth]{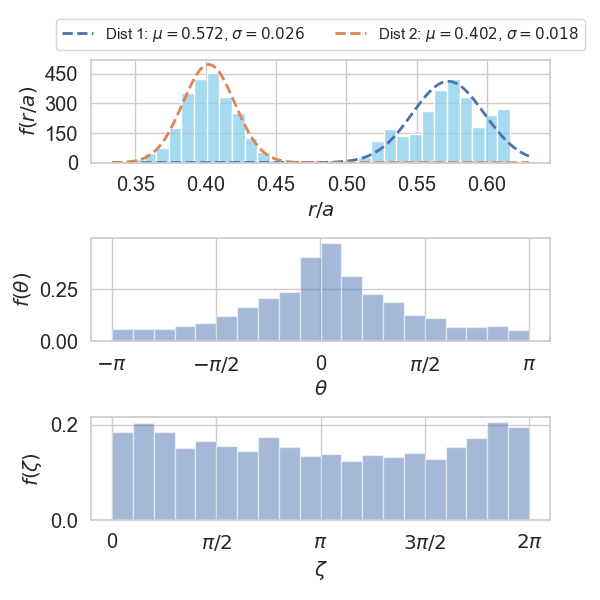}
\caption{Distribution functions with the two populations.}
\end{subfigure}
\begin{subfigure}[t]{0.45\textwidth}
\centering
\includegraphics[width=0.95\linewidth]{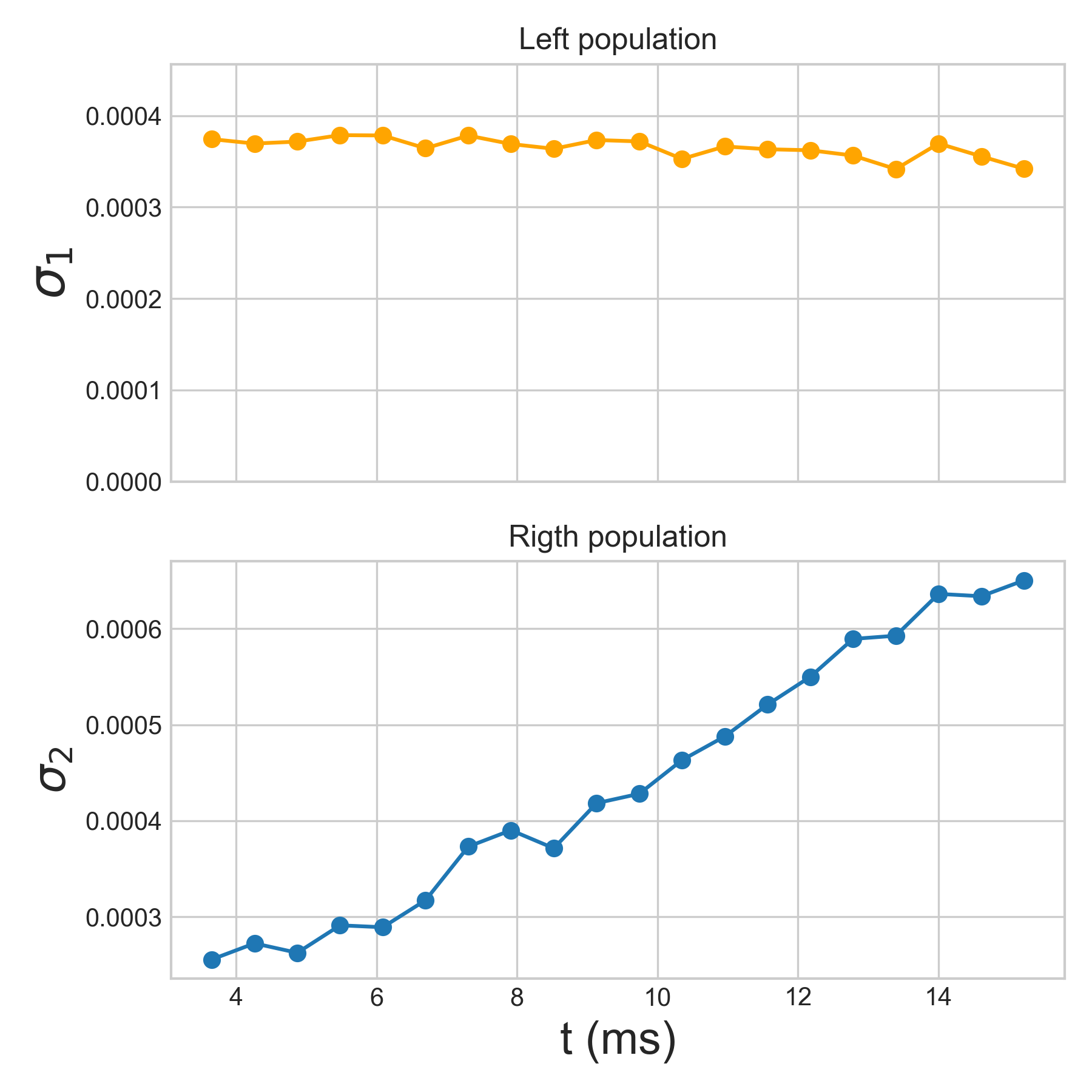}
\caption{Variance evolution.}
\end{subfigure}
\caption{Behavior of the two populations: trapped particles are associated with the left-hand population whose variance remains constant on average in time, and passing particles represented by the right-hand population which has an increasing variance in time. Here $r/a=0.4$,  $n_w=5$ and $A=1$ V.}
\label{fig:A10e5_R4_E1_n5_PT20_ISO_SEP}
\end{figure}

There were 17 configurations under the deterministic regime that exhibited this behavior. Then, the properties of each population were determined by fitting a Gaussian function model, thereby extracting the respective statistical information, in particular, the behavior of the variance of each population. Figure \ref{fig:A10e5_R4_E1_n5_PT20_ISO_SEP} shows the general behavior. Once the populations separate, they evolve independently over time. On the one hand, the distribution on the left maintains an almost constant variance, meaning that it does not broaden with time. On the other hand, the distribution on the right exhibits a linear increase in variance.

Based on the fact that under deterministic phases, pitch angles corresponding to trapped trajectories never produce transport, it is reasonable to infer that the observed separation phenomenon indeed splits the passing trajectories from the trapped ones, leaving the trapped trajectories without transport, while the passing trajectories are displaced to larger radii and exhibit a diffusive transport.

In addition, it was shown that the effect of drift waves on passing trajectories is to displace them toward radial positions farther from the magnetic axis, exactly as observed in these cases (c.f. Fig~\ref{fig:ORBITS_DRIFTS_DET}).

From these two observations, it can be concluded that the transport induced by drift waves affects the two types of trajectories differently, and it is this difference that gives rise to the observed population separation, where one population exhibits transport (the passing trajectories) while the other does not.

\section{Presence of zonal flows \label{s6}}

A common way of modifying transport in toroidal devices is by the presence of zonal flows. This is actually the origin of the H-mode in tokamaks which causes a transport barrier near the plasma edge. Here we can add a zonal flow by introducing an extra term into the potential of the form,

% \begin{equation}
%     \begin{aligned}
%         \phi_{ZF} = A_{ZF}2\psi_w &\left[  \sqrt{\frac{\psi}{\psi_w}} \tanh\left(\frac{\sqrt{\frac{\psi}{\psi_w}} -c}{\sigma_f} \right) \right.\\
% & \quad-\left.\sigma_f \log\left(\cosh\left(\frac{\sqrt{\frac{\psi}{\psi_w}} -c}{\sigma_f} \right)\right)\right]
%     \end{aligned}
% \end{equation}

% \begin{equation}
%     \begin{aligned}
%         \phi_{ZF}(r) =  A_{ZF}a^2B_0 &\left[  \frac{r}{a} \tanh\left(\frac{\frac{r}{a} -c}{\sigma_f} \right) \right.\\
% & \quad-\left.\sigma_f \log\left(\cosh\left(\frac{\frac{r}{a} -c}{\sigma_f} \right)\right)\right]
%     \end{aligned}
% \end{equation}

\begin{equation}
    \begin{aligned}
        \phi_{ZF}(r) =  A_{ZF}\sigma_f \, a \tanh\left(\frac{\frac{r}{a} -c}{\sigma_f} \right) 
    \end{aligned}
\end{equation}
where $\sigma_f$ is the radial width of the flow and it is centered at $r/a=c$. The strength of the flow is given by $A_{ZF}$.
This gives rise to a localized radial electric field which in turn produces a poloidal drift velocity in a narrow radial region ($\sigma_f\ll 1$ is assumed) which modifies the particle trajectories. By analogy to the previous works in 2D \cite{delCastilloNegrete2000chaotic, Martinell-dcn2013, tafoyaBarreras2022}, one would expect that the poloidal zonal flow reduce, or even suppress the chaos in that region, leading to the creation of a transport barrier.

In order to test this conjecture we followed the particle ensemble when the zonal flow is set in a region close to the initial position of the particles. Then the strength of the flow is increased in order to determine if and when the transport is reduced or completely suppressed. An effective transport barrier would be expected to occur only when transport is due to chaotic orbits; in that case the last surviving (to chaos) KAM surface in phase-space plays the role of the barrier, for no particle trajectory can cross it. On the other hand, the cases where transport arises because of the random phases of the waves should not present transport barriers that block the passage of particles across them. At most, the poloidal flow may slightly reduce the radial transport. With the intention of comparing the two cases, the effect of zonal flows on transport, for random and deterministic phases, was studied.

\subsection{Random phases}

\begin{figure}[hbp]
	\centering
	\begin{subfigure}[t]{0.48\columnwidth}
		\centering
		\includegraphics[width=\linewidth]{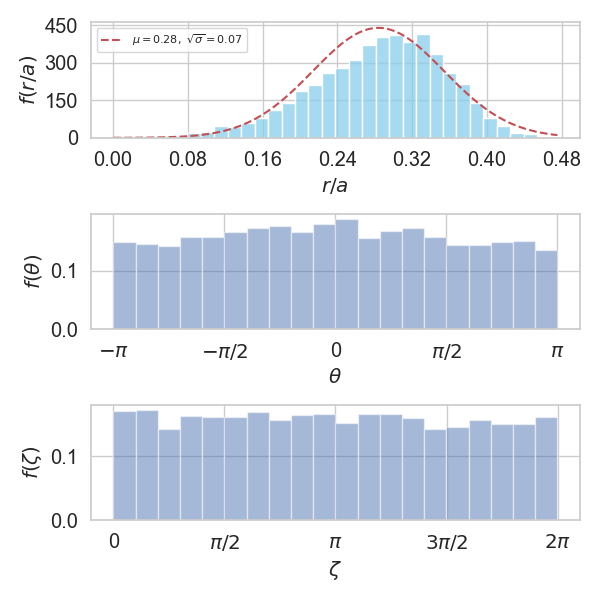}
		\caption{Final distributions without zonal flow .}
	\end{subfigure}
	\hfill
	\begin{subfigure}[t]{0.48\columnwidth}
		\centering
		\includegraphics[width=\linewidth]{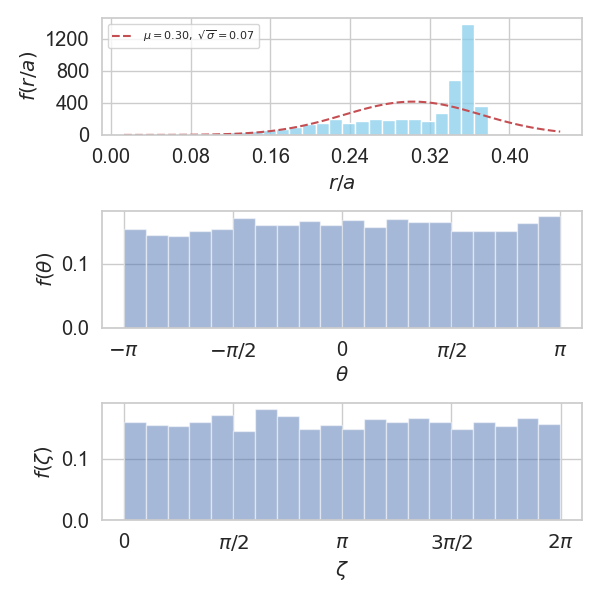}
		\caption{Final distributions with zonal flow ($A_{ZF}=1$ V).}
	\end{subfigure}

	\vspace{0.5em}

	\begin{subfigure}[t]{0.48\columnwidth}
		\centering
		\includegraphics[width=\linewidth]{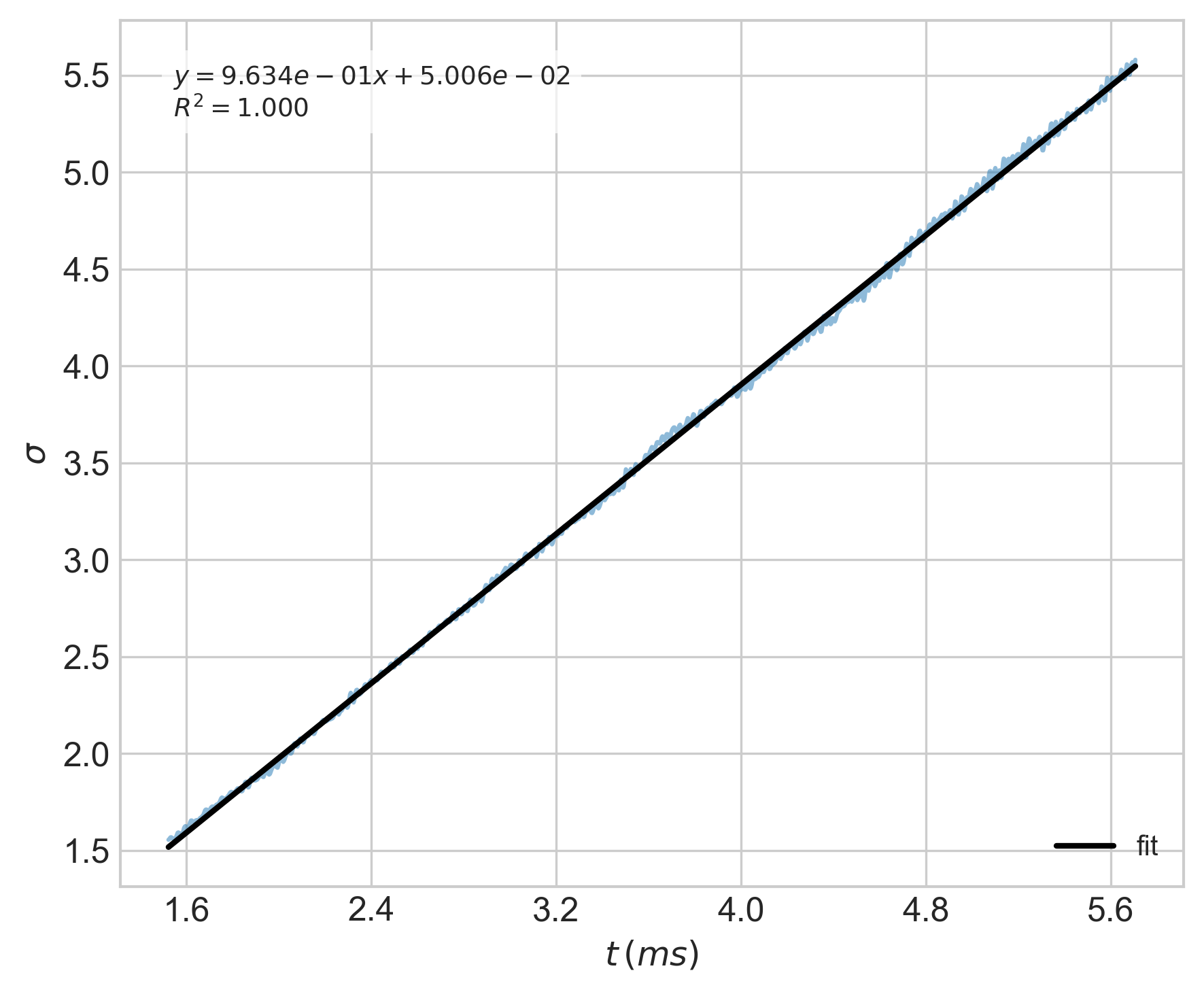}
		\caption{Variance evolution without zonal flow.}
	\end{subfigure}
	\hfill
	\begin{subfigure}[t]{0.48\columnwidth}
		\centering
		\includegraphics[width=\linewidth]{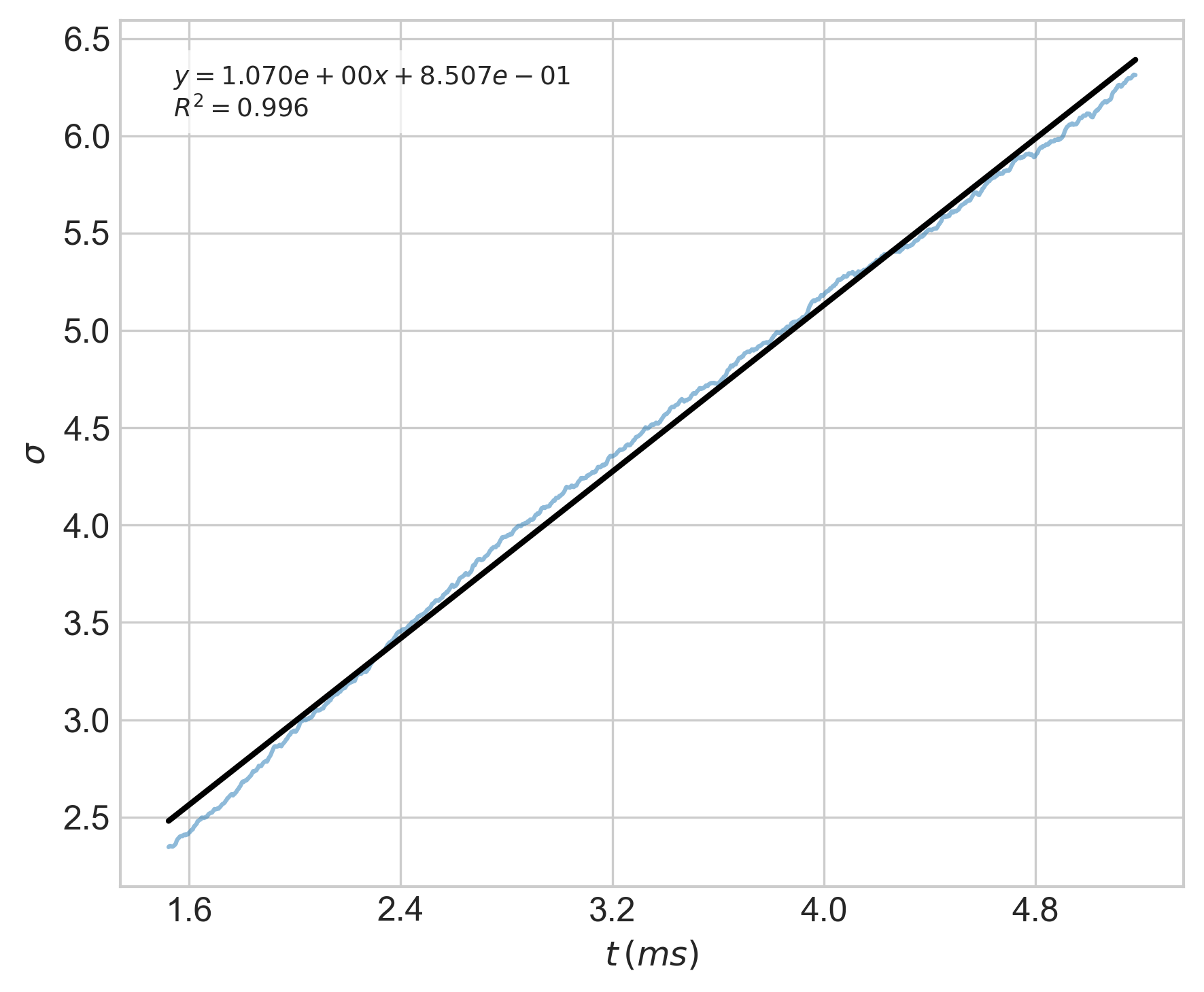}
		\caption{Variance evolution with zonal flow ($A_{ZF}=1$ V).}
	\end{subfigure}

	\caption{Effects of the zonal flow due to a fluctuation spectrum potential with random phases, $A=0.1$ V, $nw=10$, initial position at $r/a=0.3$, and a zonal flow strength of $A_{ZF}=1$ V centered at $r/a=0.31$ with $\sigma_f=  0.01$. %
    Particle distributions with no flow in (a) and with zonal flow in (b) which shows a redistribution of particles about the centroid of the flow. No significant reduction of transport results as  shown by the evolution of the variance without (c) and with zonal flow (d).}
	\label{fig:A10e4_R3_E1_n10_PT20_fix_RDM_ZF}
\end{figure}

When dealing with random phases the transport is produced by a stochastic process and hence it does not rely on the occurrence of chaos, so even in presence of a zonal flow, the transport is not expected have an important modification, although a moderate reduction could be expected. This is because the poloidal flow tends to drag the particles in the poloidal direction, thereby reducing the radial motion.
Figure \ref{fig:A10e4_R3_E1_n10_PT20_fix_RDM_ZF}  illustrates the effect of the zonal flow when transport is due to random phases in the waves. Panel (a) shows the final distributions when there is no zonal flow, which is a widening Gaussian, as already seen before. Panel (b), in contrast, shows how the distribution is affected by a zonal flow centered at $c=0.31$. It is clear that there is
a redistribution of particles on the side where the flow is located, having a strong peak at a radius larger than the center of the flow (around $r/a=0.35$). This seems to suggest that, as the particles reach the localized flow they are dragged poloidally and reduce their radial displacement, so they start accumulating at that radius. However, their stochastic motion continues and they cross to the other side of the flow region at a slower pace than in the absence of flow. This means that the poloidal flow diminishes radial transport but does not cut it off altogether. It does not constitute a full transport barrier, in the sense of blocking the transport. 

Additionally, diffusive transport, measured by the time evolution of the variance of the distribution is hardly changed by the flow, as shown in panels (c) and (d). It has to be noted, though, that the widening of the variance for a non-Gaussian function does not necessarily represent a transport process, as in the present case, where there is a separation of populations. But it still gives an estimate of the effectiveness of the radial dispersion produced by the underlying process. Then, we see that the radial electric field does not alter the radial propagation of the particles significantly. 
This behavior was observed for all flow strengths studied.

These results lead us to conclude that no transport barrier is formed when the waves have a random distribution of phases.

% \begin{figure}[htbp]
% 	\centering
% 	\begin{subfigure}[t]{0.48\textwidth}
% 		\centering
% 		\includegraphics[width=0.85\linewidth]{Imagenes/A10e4_R5_E1_n2_PT20_fix_RDM_compare/mome_compare.png}
% 		\caption{Variance evolution.}
% 	\end{subfigure}
%     	\begin{subfigure}[t]{0.48\textwidth}
% 		\centering
% 		\includegraphics[width=0.85\linewidth]{Imagenes/A10e4_R5_E1_n2_PT20_fix_RDM_compare/dist_25_compare.png}
% 		\caption{Final distribution.}
% 	\end{subfigure}

% 	\caption{ Effects due to electric potential A=0.1 V and 2 waves and zonal flow  $A_{ZF}$ = 1 V. with center $\frac{r}{a} = 0.515$ (a) Show how the evolution of the variance was affected by the zonal flow. (b) Show how the PDF was  modified by the zonal flow but still it did not form a barrier.}
% 	\label{fig:A10e4_R5_E1_n2_PT20_fix_RDM_ZF}
% \end{figure}

\subsection{Deterministic phases}

For deterministic phases, in which a chaotic regime is believed to be responsible for the transport, the zonal flow is expected to form a transport barrier, as it happens in two-dimensional models with constant magnetic field \cite{Martinell-dcn2013, tafoyaBarreras2022}. An analysis of the four cases presented in Section~\ref{s3d} that showed chaotic features was done with the inclusion of the zonal flow. The effect of the flow strength on transport and the possible appearance of a barrier is explored. For each case, the flow centroid was placed at a radius that depends on the initial position of the particle ensemble. It was chosen based on the observed distributions in the absence of flow, so that the particles would feel the effect of the flow shortly after the initial time. Then, we accordingly set $c=0.35, 0.4, 0.46, 0.53$ for the respective cases of Table \ref{tab:FD_10E5_PT20}.  

\begin{figure}[htp]
	\centering
	\begin{subfigure}[t]{0.48\columnwidth}
		\centering
		\includegraphics[width=\linewidth]{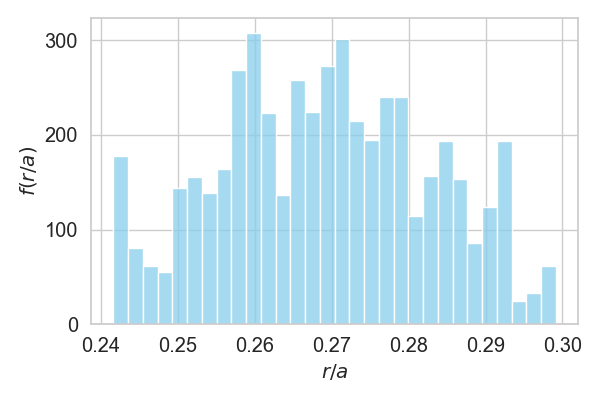}
		\caption{Radial distribution at 3 ms.}
	\end{subfigure}
	\hfill
	\begin{subfigure}[t]{0.48\columnwidth}
		\centering
		\includegraphics[width=\linewidth]{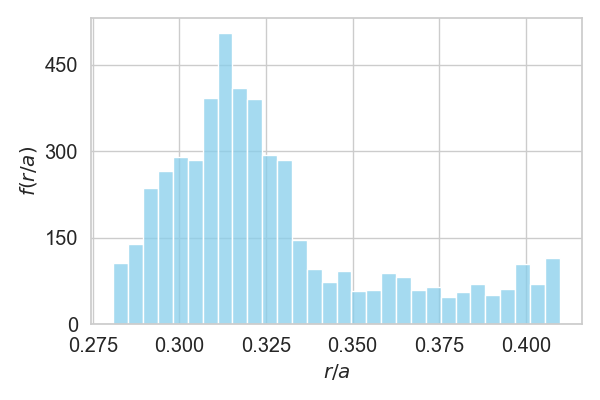}
		\caption{Radial distribution at 5.5 ms.}
	\end{subfigure}

	\vspace{0.5em}

	\begin{subfigure}[t]{0.48\columnwidth}
		\centering
		\includegraphics[width=\linewidth]{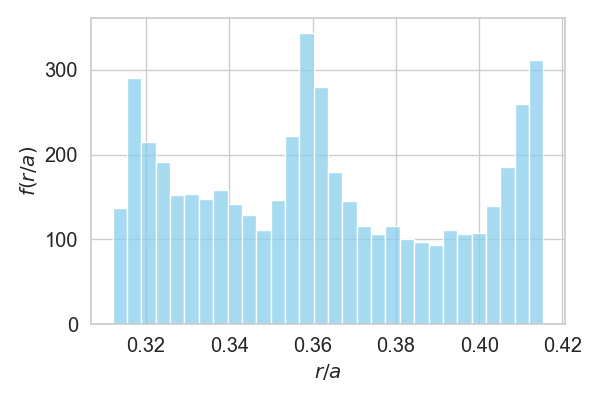}
		\caption{Final distribution at 15.2 ms}
	\end{subfigure}
	\hfill
	\begin{subfigure}[t]{0.48\columnwidth}
		\centering
		\includegraphics[width=\linewidth]{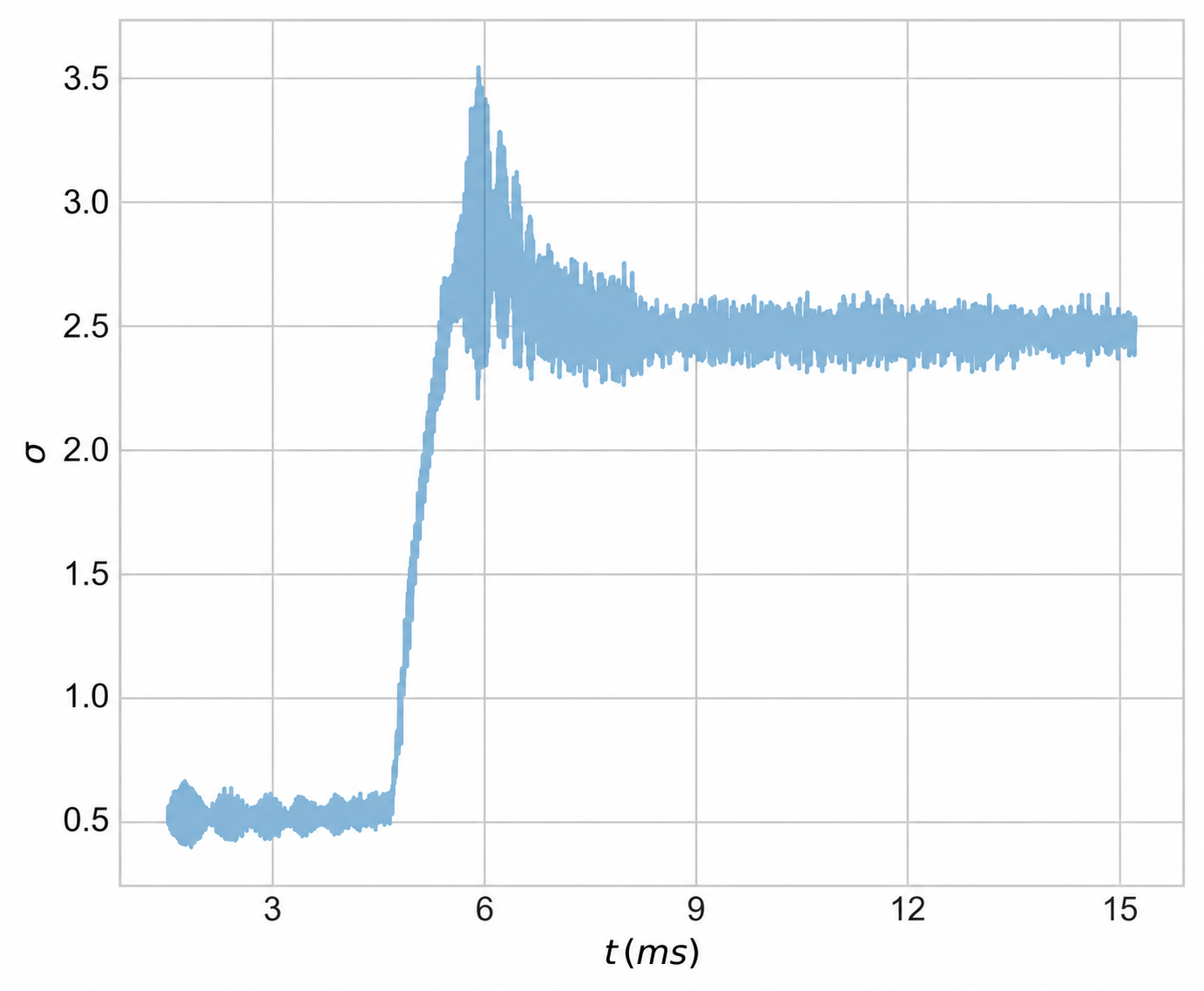}
		\caption{$A_{ZF}$ = 0.01 V}
	\end{subfigure}
    \caption{Time evolution of particle distribution and its variance for the first case of Table \ref{tab:FD_10E5_PT20} with wave potential $A= 1$ V, $nw=2$, initial radial position $r/a = 0.2$ and pitch angle $20^{\circ}$. A weak zonal flow was added with $A_{ZF}$ = 0.01V , centered at $r/a = 0.35$, and with $\sigma_f = 0.01$. (a) Shows the expansion of the distribution due to transport prior to facing the zonal flow. (b) Shows that particles affected by the flow are able to cross the maximum of the zonal flow. (c) Exhibits that there is a peak at $r/a=0.36$ due to the effects of the zonal flow but a large fraction of particles have crossed to the other side, not showing the presence of a transport barrier. (d) The evolution of the variance depicts the changes due to the zonal flow changing the width but not forming a transport barrier.}
	\label{fig:evolution_with_flow}
\end{figure}

A general behavior observed in all four cases is that, when the amplitude of zonal flow potential is very small, particles are able to cross the position of the maximum of the flow velocity, as they are transported radially outward from their initial position. This is illustrated in Figure \ref{fig:evolution_with_flow} in which the time evolution of the particle distribution is shown for the second case of Table \ref{tab:FD_10E5_PT20}, with a flow strength of $A_{ZF}=0.01$ V. Panel (a) presents the distribution shortly after initiating the simulation, in which the particles have been distributed in radius according to the diffusive transport resulting from the chaotic trajectories, just as when there is no flow. At a later time, in Panel (b) some particles have reached the region affected by the flow, which causes the distribution to widen abruptly as those particles cross to the other side of the flow quite efficiently. At the final time, in panel (c), about half of the particle population crossed to the other side of the flow but although the maximum of the distribution is at the flow position it is not Gaussian; the effect of the flow is apparently to retain the particles there, as they are dragged in poloidal direction which reduces the radial transport; this produces a piling up of the particles that reach the radial position where the flow velocity is maximum (in this case $r=0.35a$) and they steadily cross to the other side of the flow. Although the transport is slowed down overall, there is no indication of a barrier. In panel (d) it is shown the evolution of the variance, where it is apparent the sudden increase of the radial width when the particles reach the flow position, followed by a smaller variance which is kept steady. Since the distributions are not Gaussian, the variance evolution cannot be related to a diffusive transport process.

\begin{figure}[htp]
	\centering
	\begin{subfigure}[t]{0.48\columnwidth}
		\centering
		\includegraphics[width=\linewidth]{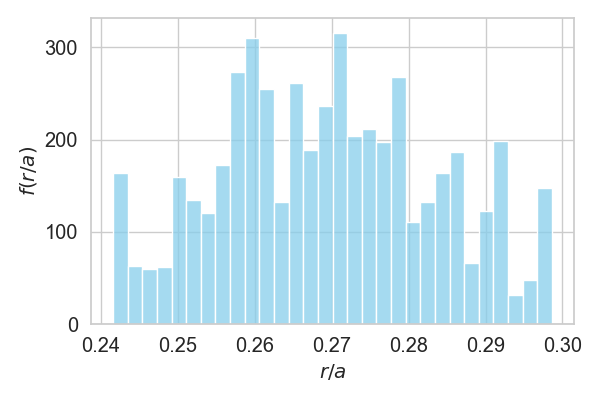}
		\caption{Radial distribution at 3 ms.}
	\end{subfigure}
	\hfill
	\begin{subfigure}[t]{0.48\columnwidth}
		\centering
		\includegraphics[width=\linewidth]{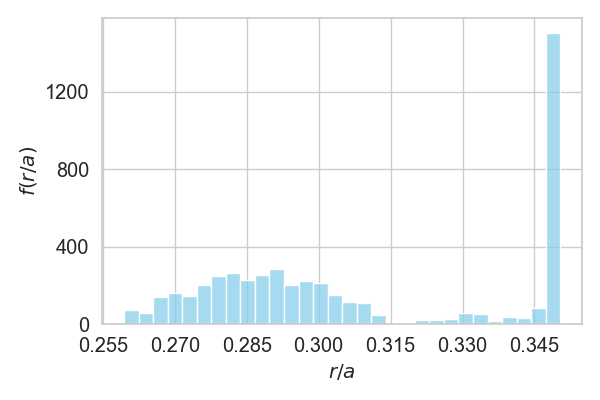}
		\caption{Radial distribution at 4.25 ms.}
	\end{subfigure}

	\vspace{0.5em}

	\begin{subfigure}[t]{0.48\columnwidth}
		\centering
		\includegraphics[width=\linewidth]{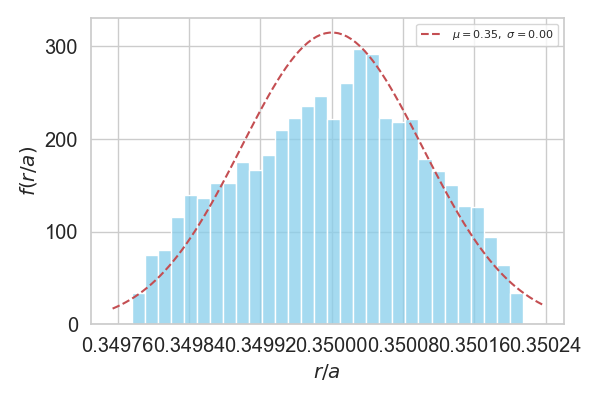}
		\caption{Final distribution at 15.2 ms}
	\end{subfigure}
	\hfill
	\begin{subfigure}[t]{0.48\columnwidth}
		\centering
		\includegraphics[width=\linewidth]{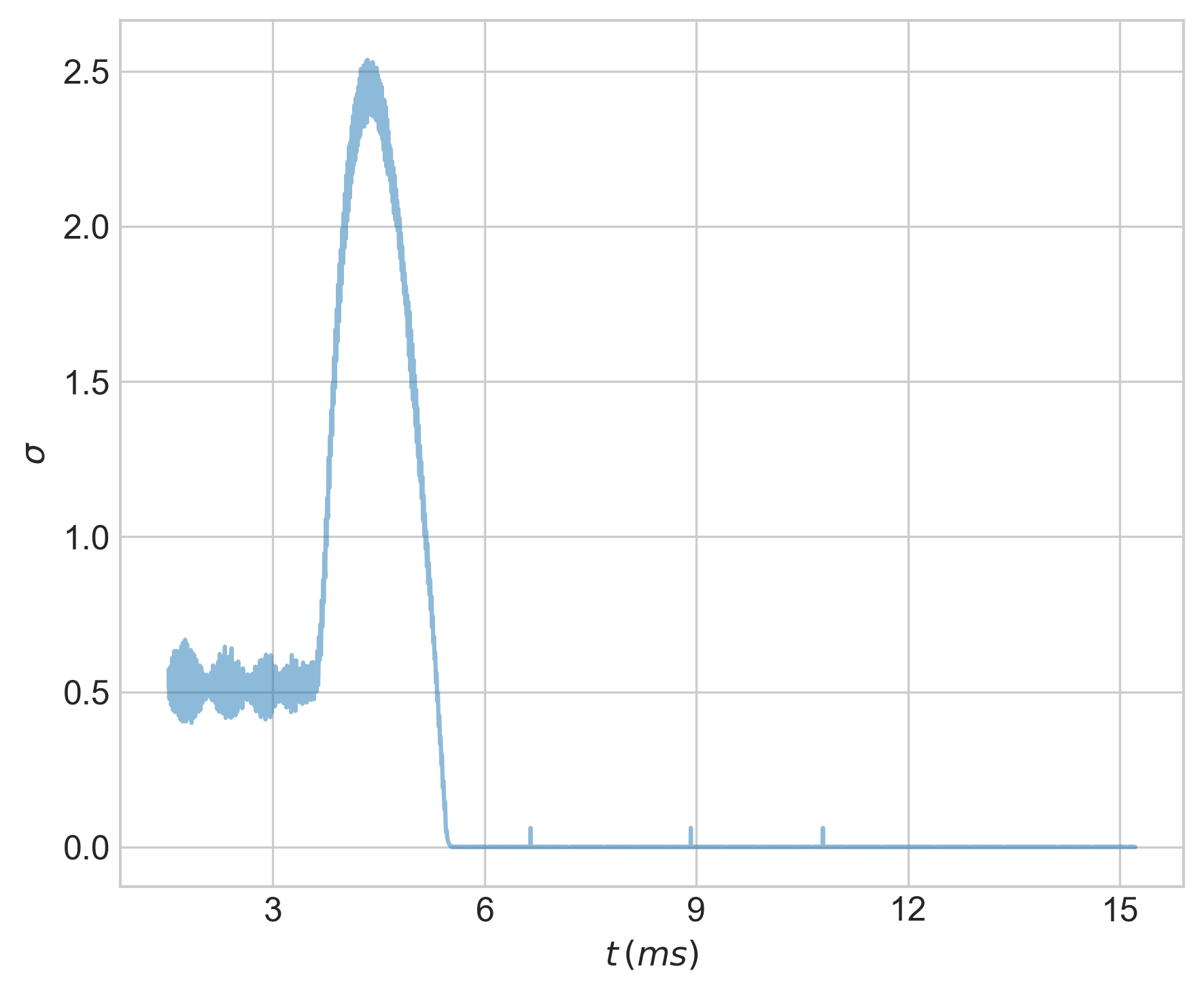}
		\caption{$A_{ZF}$ = 1 V}
	\end{subfigure}
    \caption{Same as Figure \ref{fig:evolution_with_flow} but with a stronger zonal flow with $A_{ZF}$ = 0.5V , centered at $r/a = 0.35$, and with $\sigma_f = 0.01$. (a) Shows the expansion of the particle distribution due to transport before it encounters the zonal flow. (b) Shows that particles being affected by the flow are stopped before reaching its maximum. (c) Shows that all particles have been transported to a radius corresponding to the zonal flow position and reached a gaussian distribution, indicating that a transport barrier has been formed. (d) Depicts how the zonal flow modifies the time evolution of the variance eventually leading to a nearly zero width.}
	\label{fig:evolution_with_flow_BARRIER}
\end{figure}

For larger flow speeds a different behavior shows up that indicates the appearance of a transport barrier. In Figure \ref{fig:evolution_with_flow_BARRIER} on can observe the time evolution of the distribution and of its variance for the same case of Figure \ref{fig:evolution_with_flow} but for a flow amplitude of $A_{ZF}=0.5$ V. Panel (a) shows that the radial distribution at an early time ($t=3$ ms) is still essentially the same as in Figure \ref{fig:evolution_with_flow} due to a chaotic diffusive transport, prior to encountering the flow potential. However, shortly afterwards ($t=4.25$ ms), in the distribution of panel (b), although most particles are still experiencing the usual diffusion, a small fraction has reached the surface at $r=c$ without crossing it. This produces a large widening of the distribution that gives a large value for the variance $\sigma$. This is seen in panel (d) where $\sigma(t)$ has a large increase, followed by a sharp reduction. This drop comes because the particles reaching $r/a=c$ are unable to cross and are overcome by the rest of the population which are being dragged by the flow. As a result, the final distribution shown in panel (c) has an approximate gaussian shape but with an extremely small variance (essentially zero). As panel (d) shows, the variance stays close to zero for the last part of the simulation meaning that there is no transport: a transport barrier is blocking the pass to all particles. The first part of the graph has no meaning for determining a transport process since the distributions are non-Gaussian.

The same behavior is observed for the other three cases of Table \ref{tab:FD_10E5_PT20} that we assumed to have chaotic orbits. This seems to support the thesis that chaos-induced transport can be blocked by the appearance of a transport barrier linked to the presence of a shearless surface of the flow which separates two chaotic regions of the space, in complete analogy to the 2D Hamiltonian models.

This example is for a flow strength $A_{ZF}=0.5$ V but the same trend in observed for all strengths larger than 0.01 V that were simulated. This means that there should be a threshold value for the formation of the barrier which is above 0.01 V. It is presumed that above the threshold a robust shearless surface is formed having particle regular motion and separates the regions of chaotic orbits on both sides of $r/a=c$.  However, we cannot ascertain the existence of the threshold with the method used here.

\section{Discussion \label{s7}}

Based on the results for the particle motion in the toroidal magnetic field of a tokamak under the effect of model drift waves, we can highlight the important features that point to the presence of chaotic orbits that leads to a radial particle transport. The methodology employed compares the behavior of an ensemble of particles under the action of two possible wave spectra. The first one uses random phases for the waves and as such leads naturally to a stochastic process, which in turn gives rise to diffusive transport. This is exemplified in Fig.~\ref{fig:A10e4_R5_E1_n2_PT20_fix_RDM}. For the second spectrum the phases are deterministic so that no transport would be expected when all particles follow regular orbits. Only when the trajectories become chaotic can there be transport. Since the non-integrability of the system appears just for certain circumstances and initial conditions, and our system contains several variables, the parameter scan we did could only identify four clearly cut cases showing transport, given in Table ~\ref{tab:FD_10E5_PT20}. These have a linear increase, on average, of the particle distribution variance with time, in addition to keeping an approximate Gaussian PDF, as seen in Fig.~\ref{fig:A10e5_R3_E1_n1_PT20_fix}. This behavior can be taken as evidence of chaotic diffusive transport and it is analogous to the stochastic transport produced by the random phases.

The chaos conjecture was further tested by computing the finite-time Lyapunov exponents for the cases of Table~\ref{tab:FD_10E5_PT20}. While the computation method was only approximate since it did not follow the particles to the asymptotically infinite time, the values obtained are consistent with chaotic motion: they are all positive and in the magnitude range of the simulation times (milliseconds) as shown in Table~\ref{tab:FTLE}. This means that orbits having very similar initial conditions diverge exponentially in times of ms, for the conditions suspected to present chaos. In contrast, all the cases with random phases had a very small Lyapunov exponent, which gives almost no divergence for times in the ms range. This comparison reinforces the hypothesis of chaotic orbits.

Then we focused on the important feature observed in Hamiltonian systems that model particle motion in 2D, when zonal flows are present, namely, the presence of transport barriers when they are perturbed by time-dependent fluctuations \cite{del-castillo93,delCastilloNegrete2000chaotic,Martinell-dcn2013}.
The streamlines of the flow in the 2D configuration space are identified with KAM torii in the phase-space of the equivalent 1D Hamiltonian system, so when chaos is developed the torii are broken sequentially starting with those having rational winding numbers. Since zonal flows have non-monotonic variation with radius they correspond to a non-twist Hamiltonian, for which there is a robust torus that remains unbroken corresponding to the shearless surface (maximum of the flow velocity). This acts as a transport barrier preventing particles from crossing radially. Eventually, the shearless torus is destroyed and global chaos sets in, allowing particles to be transported all over the domain. Although we cannot directly observe this behavior in the 3D simulations, we can test the appearance of a transport barrier with the mentioned properties, if a zonal flow is added to our model. Doing this, we found that, in the case of random phases, there is not a real barrier but just a particle redistribution in radius, for all flow velocity magnitudes, as seen in Fig.~\ref{fig:A10e4_R3_E1_n10_PT20_fix_RDM_ZF}. This is what would be expected in a non-chaotic system. However, for deterministic phases under situations where diffusive transport was observed, particle distribution was significantly modified for flow amplitude larger than $A_{ZF}=0.01$ V. For all the cases in Table \ref{tab:FD_10E5_PT20} it was observed that no particles can cross the zonal flow centroid, as seen in Fig.~\ref{fig:evolution_with_flow_BARRIER}. This is interpreted as though global chaos is present when there is no zonal flow, but when the flow is set and exceeds a threshold strength, a shearless surface KAM torus appears, preventing particles to go through. This observation then gives further support to the chaotic origin of transport.

Putting together all the results, there is good evidence that the origin of transport in the cases with no randomness present, is due to the appearance of chaotic orbits, which are expected to occur in our time-dependent Hamiltonian system. 

It is worth mentioning that this work did not explore the intermediate regime consisting of a mixed phase space, in which regular structures coexist with a surrounding chaotic sea, giving rise to the well-known phenomenon of sticky orbits \cite{souza2024}. Such behavior could nonetheless be inferred indirectly by characteristic changes in both the PDF and the evolution of the variance, where such phenomena would produce a nonlinear growth of the variance together with a non-gaussian distribution, a bimodal distribution produced by the interference of particles trapped near regular structures and the ones following chaotic trajectories.

\section{Conclusions \label{s8}}

The simulations presented here have shown important differences in the transport properties derived from the two models of electrostatic drift waves explored, namely, random and deterministic phases. The purpose of the study was to show that chaotic particle orbits can arise in the Hamiltonian system describing the charged particles three-dimensional motion in the toroidal magnetic field of a tokamak, in presence of a spectrum of electrostatic waves. One of the models considers the phases of the waves given by a random distribution, giving an intrinsic stochastic process. In contrast, the second model has deterministic phases but there is the possibility of having chaotic trajectories, due the time dependence of the Hamiltonian.

The results showed that diffusive transport is always present in the cases with random phases, since an underlying stochastic process is causing it. When deterministic phases are used only a few cases show transport and the cause can be traced to the appearance of chaotic particle orbits. Three pieces of evidence were given to support this assertion. The first is simply the fact that, since the properties associated with diffusive transport are observed (distribution function broadening with linear increase in time) in the deterministic model, the only possible cause should be the presence of chaos. The second refers to the more quantitative characterization in terms of Lyapunov exponents, $\lambda$. The computation of $\lambda$ was consistent with the existence of chaos, for initially neighboring orbits diverge exponentially in the times involved in the simulations. The third evidence comes from the addition of a zonal flow, which is linked to the appearance of a transport barrier. It was observed that, above a threshold value of the flow velocity, a clear transport barrier is formed that completely blocks the radial crossing of particles. This is in accordance with the behavior of non-twist Hamiltonian systems describing the 2D motion of magnetized plasma particles. 

\

\leftline{\Large\bf Acknowledgments}

\

Useful discussions with D. del-Castillo-Negrete on some of the topics of the manuscript are greatly acknowledged. The work was partially supported by DGAPA-UNAM PAPIIT project IN105125.

\bibliographystyle{unsrt}%{abbrv}
\bibliography{transp}

@article{horton99,
    author  = "W. Horton",
    title   = "Drift waves and transport",
    year    = "1999",
    journal = "Reviews of Modern Physics",
    volume  = "71",
    number  = "3",
    pages   = "735--778"
}

@article{maeyama24,
    author  = "S. Maeyama and T. Tokuzawa and N.T. Howard and J. Citrin and T.-H. Watanabe",
    title   = "Overview of multiscale turbulence studies covering ion-to-electron scales in magnetically confined fusion plasma",
    year    = "2024",
    journal = "Nuclear Fusion",
    volume  = "64",
    number  = "11",
    pages   = "112007"
}

@article{del-castillo93,
    author  = "D. del-Castillo-Negrete and P.J. Morrison",
    title   = "Chaotic transport by Rossby waves in shear flow",
    year    = "1993",
    journal = "Phys. Fluids A",
    volume  = "5",
    number  = "4",
    pages   = "948--965",
	doi     = {10.1063/1.858639}
}

@article{kleva84,
    author  = "R.G. Kleva and J.F Drake",
    title   = "Stochastic EXB particle transport",
    year    = "1984",
    journal = "Phys. Fluids",
    volume  = "27",
    number  = "7",
    pages   = "1686--1698",
	doi     = {10.1063/1.864823}
}

@article{burrell97,
    author  = "K.H. Burrell",
    title   = "Effect of {EXB} velocity shear and magnetic shear on turbulence and transport in magnetic confinement devices",
    year    = "1997",
    journal = "Physics of Plasmas",
    volume  = "4",
    number  = "5",
    pages   = "1499--1518",
	doi       = "10.1063/1.872367"
}

@book{white2014,
	author    = {White, R. B.},
	title     = {The theory of toroidally confined plasmas},
	publisher = {Imperial College Press},
	address   = {London},
	year      = {2014}
}

@book{wessonTokamaks2004,
	author    = {Wesson, J. A.},
	title     = {Tokamaks},
	publisher = {Clarendon Press},
	edition   = {3rd},
	address   = {Oxford},
	year      = {2004}
}

@article{ferro_caldas18,
	author    = {Ferro, R.M. and Caldas, I.L.},
	title     = {Internal transport barriers in plasmas with reversed plasma flow},
	journal   = {Physics Letters A},
	volume    = {382},
	number    = {15},
	pages     = {1014--1019},
	year      = {2018},
	doi       = {10.1016/j.physleta.2018.02.019}
}

@article{delcastillo96,
	author    = {del-Castillo-Negrete, D. and Greene, J.M. and Morrison, P.J.},
	title     = {Area preserving nontwist maps: periodic orbits and transition to chaos},
	journal   = {Physica D: Nonlinear Phenomena},
	volume    = {91},
	number    = {1--2},
	pages     = {1--23},
	year      = {1996},
	doi       = {10.1016/0167-2789(95)00257-X}
}

@article{delCastilloNegrete2000chaotic,
	author       = {del-Castillo-Negrete, Diego},
	title        = {Chaotic transport in zonal flows in analogous geophysical and plasma systems},
	journal      = {Physics of Plasmas},
	volume       = {7},
	number       = {5},
	pages        = {1702--1711},
	year         = {2000},
	doi          = {10.1063/1.873988},
	note         = {41st Annual Meeting of the Division of Plasma Physics of the American Physical Society (Seattle, WA, 1999)} 
}

@article{morrison2000,
	author    = {Morrison, P.J.},
	title     = {Magnetic field lines, Hamiltonian dynamics, and nontwist systems},
	journal   = {Physics of Plasmas},
	volume    = {7},
	number    = {9},
	pages     = {2279--2289},
	year      = {2000},
	doi       = {}
}

@article{kryukov2018,
	author    = {Kryukov, N. and Martinell, J. J. and del-Castillo-Negrete, D.},
	title     = {Finite Larmor radius effects on weak turbulence transport},
	journal   = {Journal of Plasma Physics},
	volume    = {84},
	number    = {3},
	year      = {2018},
	doi       = {10.1017/s0022377818000351}
}

@article{Martinell-dcn2013,
	author    = {Martinell, J. J. and del-Castillo-Negrete, D.},
	title     = {Gyroaverage effects on chaotic transport by Drift Waves in zonal flows},
	journal   = {Physics of Plasmas},
	volume    = {20},
	number    = {022303},
	year      = {2013},
	doi       = {10.1063/1.4790639}
}

@article{delcastillo-Martinell2012,
	author    = {del-Castillo-Negrete, D. and Martinell, J. J.},
	title     = {Gyroaverage effects on nontwist Hamiltonians: Separatrix reconnection and chaos suppression},
	journal   = {Communications in Nonlinear Science and Numerical Simulations},
	volume    = {17},
	pages    = {2031},
	year      = {2012}
}

@article{tafoyaBarreras2022,
	author    = {Tafoya, C. A. and Martinell, J. J.},
	title     = {Breakup of transport barriers in plasmas with flow described by symplectic maps},
	journal   = {Radiation Effects and Defects in Solids},
	volume    = {177},
	number    = {1--2},
	pages     = {124--136},
	year      = {2022},
	doi       = {10.1080/10420150.2022.2049787}
}

@article{torresTransporteTurbulento2023,
	author    = {Torres, J. and Martinell, J. J.},
	title     = {Study of turbulent transport in magnetized plasmas with flow using symplectic maps},
	journal   = {Chaos: An Interdisciplinary Journal of Nonlinear Science},
	volume    = {33},
	number    = {5},
	year      = {2023},
	doi       = {10.1063/5.0144037}
}

@book{Chaos_lichtenberg_lieberman_1983,
	author    = {Allan J. Lichtenberg and Michael A. Lieberman},
	title     = {Regular and Stochastic Motion},
	year      = {1983},
	publisher = {Springer-Verlag},
	address   = {New York},
	edition   = {1st},
	isbn      = {978-0-387-90885-5}
}

@article{Finn-delCastillo2001,
	author       = {Finn, John M. and del-Castillo-Negrete, Diego},
	title        = {Lagrangian chaos and Eulerian chaos in shear flow dynamics},
	journal      = {Chaos},
	volume       = {11},
	number       = {4},
	pages        = {816--832},
	year         = {2001},
	doi          = {10.1063/1.1418762}
}

@article{macha23,
	author       = {Macha, P. and Adamek, J. and Seidl, J. and Stockel, J. and Svoboda, V. and Van Oost, G. and Lobko, L. and Krbec, J.},
	title        = {Spontaneous formation of a transport barrier in helium plasma in a tokamak with circular configuration},
	journal      = {Nuclear Fusion},
	volume       = {63},
	number       = {10},
	pages        = {104003},
	year         = {2023},
	doi          = {10.1088/1741-4326/acf1af},
}

@article{kobayashi20,
	author       = {Kobayashi, T.},
	title        = {The physics of the mean and oscillating radial electric field in the L–H transition: the driving nature and turbulent transport suppression mechanism},
	journal      = {Nuclear Fusion},
	volume       = {60},
	number       = {9},
	pages        = {095001},
	year         = {2020},
	doi          = {10.1088/1741-4326/ab7a67},
}

@book{strogatz2014nonlinear,
	title     = {Nonlinear Dynamics and Chaos: With Applications to Physics, Biology, Chemistry, and Engineering},
	author    = {Strogatz, Steven H.},
	edition   = {2},
	year      = {2014},
	publisher = {Westview Press},
	address   = {Boulder, CO},
	isbn      = {978-0-8133-4910-7}
}

@article{souza2024,
	author    = "{L.C. de Souza} and M.R. Sales and M. Mugnaine and J.D. Szezeche and I.L. Caldas and R. Viana",
	title     = {Chaotic escape of impurities and sticky orbits in toroidal plasmas},
	journal   = {Physical Review E},
	volume    = {109},
	number    = {1},
	pages     = {15202},
	year      = {2024},
	doi       = {10.1103/PhysRevE.109.015202},
	url       = {https://doi.org/10.1063/5.0266608}
}

\end{document}